\documentclass[twocolumn]{aastex63}

\pdfoutput=1

\let\tablenum\relax
\usepackage{siunitx}
\DeclareSIUnit\year{yr}
\usepackage[version=4]{mhchem}

\usepackage{array,etoolbox}
\preto\tabular{\setcounter{magicrownumbers}{0}}
\newcounter{magicrownumbers}

\usepackage{multirow}
\usepackage{makecell}

\received{March 18, 2022}
\revised{July 5, 2022}
\accepted{July 22, 2022}

\shortauthors{Sergeev et al.}

\begin{document}
\title{Bistability of the atmospheric circulation on TRAPPIST-1e}

\correspondingauthor{Denis E. Sergeev}
\email{d.sergeev@exeter.ac.uk}

\author[0000-0001-8832-5288]{Denis E. Sergeev}
\affiliation{Department of Astrophysics, College of Engineering, Mathematics, and Physical Sciences, University of Exeter, Exeter, EX4 4QL, UK}
\author[0000-0002-3724-5728]{Neil T. Lewis}
\affiliation{Atmospheric, Oceanic and Planetary Physics, Clarendon Laboratory, University of Oxford, Oxford, OX1 3PU, UK}
\author[0000-0002-4664-1327]{F. Hugo Lambert}
\affiliation{Department of Mathematics, College of Engineering, Mathematics, and Physical Sciences, University of Exeter, Exeter, EX4 4QF, UK}
\author[0000-0001-6707-4563]{Nathan J. Mayne}
\affiliation{Department of Astrophysics, College of Engineering, Mathematics, and Physical Sciences, University of Exeter, Exeter, EX4 4QL, UK}
\author[0000-0002-1485-4475]{Ian A. Boutle}
\affiliation{Met Office, FitzRoy Road, Exeter, EX1 3PB, UK}
\affiliation{Department of Astrophysics, College of Engineering, Mathematics, and Physical Sciences, University of Exeter, Exeter, EX4 4QL, UK}
\author[0000-0003-4402-6811]{James Manners}
\affiliation{Met Office, FitzRoy Road, Exeter, EX1 3PB, UK}
\author[0000-0003-0165-4885]{Krisztian Kohary}
\affiliation{Department of Astrophysics, College of Engineering, Mathematics, and Physical Sciences, University of Exeter, Exeter, EX4 4QL, UK}

\begin{abstract}
Using a 3D general circulation model, we demonstrate that a confirmed rocky exoplanet and a primary observational target, TRAPPIST-1e presents an interesting case of climate bistability.
We find that the atmospheric circulation on TRAPPIST-1e can exist in two distinct regimes for a 1~bar nitrogen-dominated atmosphere.
One is characterized by a single strong equatorial prograde jet and a large day-night temperature difference; the other is characterized by a pair of mid-latitude prograde jets and a relatively small day-night contrast.
The circulation regime appears to be highly sensitive to the model setup, including initial and surface boundary conditions, as well as physical parameterizations of convection and cloud radiative effects.
We focus on the emergence of the atmospheric circulation during the early stages of simulations and show that the regime bistability is associated with a delicate balance between the zonally asymmetric heating, mean overturning circulation, and mid-latitude baroclinic instability.
The relative strength of these processes places the GCM simulations on different branches of the evolution of atmospheric dynamics.
The resulting steady states of the two regimes have consistent differences in the amount of water content and clouds, affecting the water absorption bands as well as the continuum level in the transmission spectrum, although they are too small to be detected with current technology.
Nevertheless, this regime bistability affects the surface temperature, especially on the night side of the planet, and presents an interesting case for understanding atmospheric dynamics and highlights uncertainty in 3D GCM results, motivating more multi-model studies.
\end{abstract}

\section{Introduction}
\label{sec:intro}
The era of atmospheric characterization of rocky exoplanets is imminent with the advent of new telescopes, such as the James Webb Space Telescope (JWST, successfully launched in December 2021), the European Extremely Large Telescope (E-ELT), the Large Ultraviolet/Optical/Infrared Surveyor \citep[LUVOIR,][]{Roberge18_luvoir} and Atmospheric Remote-sensing Infrared Exoplanet Large-survey \citep[ARIEL,][]{Tinetti18_ariel}.
This study is motivated by the need to understand the atmospheric circulation on tidally locked exoplanets in order to make the best use of observational data by enabling the community to both refine target selection for observational campaigns and improve confidence in the interpretation of the observations.
Recent simulations of possible climates on TRAPPIST-1e, a rocky planet orbiting an ultracool M-dwarf star, allude to a potential bistability of the atmospheric circulation for this planet \citep[e.g.][]{Sergeev20_atmospheric,Eager20_implications,Turbet22_thai,Sergeev22_thai}.
Here we study the emergence and maintenance of two different circulation regimes of TRAPPIST-1e, a primary observational target, using a 3D general circulation model (GCM).

GCMs help us understand the variety of processes driving planetary climates.
They can reconstruct a simulated three-dimensional state of the atmosphere and its evolution, constrained by a set of parameters observed or assumed for a certain planet.
For a given planetary and atmospheric configuration, we may then obtain a long-term set of statistics (i.e. the climate) compatible with the system of equations of the numerical model.
However, multiple statistically steady solutions may be obtained for the same set of external parameters \citep{Lorenz70_climatic}.
Regions in parameter space where multiple solutions can occur are called bifurcations \citep{SuarezDuffy92_terrestrial,Saravanan93_equatorial} or bistability \citep[e.g.][]{Arnold12_abrupt,Herbert20_atmospheric}.
Moreover, there always exists an uncertainty in GCM parameters, and this is acutely felt in theoretical studies of exoplanetary atmospheres due to the extreme paucity of observational data.
This demands exploration of the model behavior over a range of parameters and configurations, alongside model intercomparisons \citep{Politchouk14_intercomparison,Yang19_simulations,Fauchez21_workshop}.

Earlier studies, focused mostly on non-tidally locked planets, discovered circulation bistability in different scenarios and in models of various degrees of complexity.
One example of circulation bistability concerns the transition to equatorial superrotation in idealized two-layer models of the Earth's atmospheric circulation (\citealp{SuarezDuffy92_terrestrial}; \citealp{Saravanan93_equatorial}; see also discussion in \citealp{Held99_equatorial}). 
Here, equatorial superrotation describes a phenomenon where the zonal wind has an excess of angular momentum relative to a state of solid body co-rotation with the underlying planet \citep{Read86_superrotation2,Read18_superrotation}, which can only be obtained when non-axisymmetric disturbances (eddies) transport angular momentum up-gradient \citep{Hide69_dynamics,Gierasch75_meridional,Rossow79_large-scale,Mitchell10_transition}.
For a fast-rotating Earth-like planet, \citet{SuarezDuffy92_terrestrial} and \citet{Saravanan93_equatorial} showed using an idealized two-layer climate model that transient eddies are affected by the strength of a heating perturbation localized at the equator.
The behavior of transient eddies changes the momentum flux balance and leads to the regime transition between a ``conventional'' state, similar to the atmospheric circulation observed on Earth, and a ``superrotating'' state, characterized by a strong eastward jet at the equator.
A positive feedback mechanism was proposed by \citet{Arnold12_abrupt} to explain the bifurcation of the circulation into a subrotating or superrotating state: the resonance of equatorial Rossby waves and background mean flow.
This mechanism was further explored by \citet{Herbert20_atmospheric} who used a simple model to prove that the wave-jet resonance is more robust relative to other feedback mechanisms suggested for the regime bistability.

In the context of tidally locked exoplanets, the bistability of atmospheric circulation was explored by \citet{Thrastarson10_effects}, \citet{Liu13_atmospheric} and \citet{Showman15_three-dimensional} for hot Jupiters, and by \citet{Edson11_atmospheric} and \citet{Noda17_circulation} for terrestrial planets.
\citet{Edson11_atmospheric} found that abrupt transitions occur between two different circulation states, with weak and strong superrotation, at a rotation period of 4--5 days for a dry planet and 3--4 days for an aquaplanet orbiting low-mass stars.
Using an idealized model with no clouds and gray radiation, \citet{Noda17_circulation} mapped the dependence of the large-scale dynamics on a range of rotation periods and identified four circulation patterns, each characterized by either thermally direct day-night circulation, wave-jet resonance, north-south asymmetric effects, or a pair of mid-latitude eastward jets.
Further research of different atmospheric regimes for abstract exoplanetary configurations was conducted by \citet{Carone14_connecting,Carone15_connecting,Carone16_connecting} and later \citet{Kopparapu17_habitable} and \citet{Haqq-Misra18_demarcating}.
The key and sometimes the only parameter demarcating the atmospheric regimes in these studies was the rotation rate of the planet.
With the number of confirmed rocky exoplanets growing, it is pertinent to explore using a 3D GCM whether a specific exoplanet can exhibit regime bistability.

For climate simulations of TRAPPIST-1e, the circulation regime can be sensitive to the representation of convection in the model, as was first noted by \citet{Sergeev20_atmospheric}.
Similar differences in circulation regimes was reported in the TRAPPIST-1 Habitable Atmosphere Intercomparison (THAI), where four different GCMs were used to simulate \ce{N2} and \ce{CO2}-dominated atmospheres of TRAPPIST-1e \citep{Turbet22_thai,Sergeev22_thai}.
The THAI project highlights that the simulated circulation regime for this planet is sensitive to the parameterizations of subgrid-scale processes in GCMs, such as boundary layer processes, radiative transfer, and moist physics.
However, the dynamical feedbacks resulting in different circulation regimes for the same planet were not explored in detail by these studies.
Providing an explanation for them will strengthen our confidence in results from current and future GCM studies, and is the main aim of this study.

In this paper, we study the two distinct circulation regimes that emerge in the atmosphere of a moist nitrogen-dominated atmosphere of TRAPPIST-1e as simulated by a 3D GCM (Sec.~\ref{sec:results_sens}).
We argue that the regime bistability originates during the model spin-up due to the different amount of water vapor in the substellar region and thus different radiative forcing, which affects the emergence of superrotation and the overall climate.
We conduct a series of experiments to test the sensitivity of the regimes to such parts of the model configuration as the initial temperature, slab ocean depth, convection scheme and the cloud radiative effect.
While the model setup is relatively idealized, with a uniform ocean surface at the lower boundary; we use the observed values of planetary radius, rotation rate and insolation with the assumption of a 1:1 synchronous rotation \citep{Grimm18_nature,Fauchez20_thai_protocol}.
We diagnose the regime evolution in the early stages of numerical simulations using various metrics of superrotation and accompanied shifts in the global climate, further supporting our arguments by looking at various terms in the angular momentum budget (Sec.~\ref{sec:results_spinup_dynamics}).
We also describe the surface conditions on the night side of the planet, because the regime shift substantially affects the night side temperature and humidity (Sec.~\ref{sec:results_spinup_cold_traps}).
Finally, we present the steady state of both regimes, showing that they are well-defined with respect to various climate diagnostics (Sec.~\ref{sec:results_mean_state}).
This has consequences for their respective imprint in the atmospheric transmission depth, mostly in the water absorption bands and the continuum level (Sec.~\ref{sec:results_synthobs}).
However, the inter-regime differences in the transmission depth are too small to be detected with the current generation of telescopes.

\section{Methodology}
\label{sec:method}
Transitions of the circulation regime for tidally locked rocky exoplanets were reported in several modeling studies, all based on 3D GCMs \citep[e.g.][]{Edson11_atmospheric,Carone14_connecting,Noda17_circulation}.
It was also noted in recent studies of the TRAPPIST-1e climate \citep{Sergeev20_atmospheric,Eager20_implications,Turbet22_thai,Sergeev22_thai}, in which the dominant climatic feature was either a strong superrotating jet at the equator or two eastward jets in mid-latitudes with a weaker equatorial superrotation.
Throughout this paper, we refer to the former as the ``single jet'' (SJ) regime, and the latter as the ``double jet'' (DJ) regime.

The analysis of the full complexity of this regime bistability, requires a 3D GCM and we employ the Met Office Unified Model (UM). 
As we report below (Sec.~\ref{sec:results}), we are able to capture both of the circulation regimes.
The control, or \emph{Base}, experiment develops the SJ regime, whilst various sensitivity simulations settle on either of the two regimes, SJ or DJ, with no intermediate states between them.
We thus choose one of the sensitivity experiments that settled on the DJ regime and compare its evolution and mean state to the SJ regime obtained in the \emph{Base} experiment.
This sensitivity experiment of choice is named \emph{T0\_280} and the only difference in its configuration to that of \emph{Base} is the initial temperature, as described in Sec.~\ref{sec:method_sens}.
Focusing on the \emph{T0\_280} case allows us to explore the bifurcation of the early stages of the simulation in a clearer way, eliminating the effect of e.g. the change of a model parameterization.
At the same time, experiments with different setups that also develop the DJ regime have similar early model evolution as well as the resulting climate, thus making our conclusions robust.

Within this section, the overall model configuration, including the planetary parameters and atmospheric composition, is described in Sec.~\ref{sec:method_model}.
Details of the setup for the base and sensitivity experiments are given in Sec.~\ref{sec:method_sens}.
The method of computing synthetic transmission spectra in our 3D simulations is also detailed in Sec.~\ref{sec:method_synthobs}.

\subsection{Model setup}
\label{sec:method_model}
\begin{deluxetable*}{lll}
\tablecaption{Stellar spectrum and planetary parameters used in this study following \citep{Fauchez20_thai_protocol}. \label{tab:planet}}
\tablehead{
\colhead{Parameter} & \colhead{Units} & \colhead{Value}
}
\startdata
Star and spectrum &                                & \SI{2600}{\K} BT-Settl with Fe/H=0 \\
Semi-major axis   & AU                             & 0.02928 \\
Orbital period    & Earth day                      & 6.1 \\
Rotation period   & Earth day                      & 6.1 \\
Obliquity         &                                & 0 \\
Eccentricity      &                                & 0 \\
Instellation      & \si{\watt\per\square\meter}    & 900.0 \\
Planet radius     & \si{\km}                       & 5798 \\
Gravity           & \si{\meter\per\second\squared} & 9.12 \\
\enddata
\end{deluxetable*}

All simulations in this study are performed with the UM (code version \texttt{vn12.0}) in the GA7.0 science configuration \citep{Walters19_ga7}.
The UM is configured at a horizontal grid spacing of \ang{2.5} in longitude and \ang{2} in latitude, with 38 vertical levels between the surface and the model top, located at a height of \SI{\approx 80}{\km}.
\footnote{We conducted an additional series of experiments with 60 and 70 vertical levels and different model top heights.
Qualitatively, the conclusions of our study are not affected: both circulation regimes emerge at a higher vertical resolution, although their dependence on the sensitivity parameters (see Sec.~\ref{sec:method_sens} below) does not exactly match those obtained in the model with 38 levels or a different model top height.}
The model is run for 3000 Earth days ($\approx$491 TRAPPIST-1e orbits) to ensure that the atmosphere reaches thermal equilibrium, when the net absorbed stellar radiation is approximately equal to the emitted thermal radiation and when the global mean surface temperature does not have a noticeable long-term trend.
Hereafter we use the word ``day'' to refer to an Earth day, i.e. \SI{86400}{\s}.
In the analysis below (Sec.~\ref{sec:results}), we use daily mean output at high temporal resolution (every day) during the spin-up phase (first 500 days) to capture the emergence of circulation patterns.
The mean-climate state is presented as the average over the days 2000--3000 of simulations (i.e. over $\approx$163 orbits).

We employ a nitrogen-dominated atmospheric configuration, used in the TRAPPIST Habitable Atmosphere Intercomparison (THAI) under the label \textit{Hab~1} \citep{Fauchez20_thai_protocol,Sergeev22_thai}, as well as in many previous exoplanet modeling studies \citep[e.g.][]{Turbet16_habitability,Wolf17_assessing,DelGenio19_albedos,Yang19_simulations}.
Namely, an atmosphere with a total mean pressure of \SI{e5}{\pascal} consisting of \ce{N2}, 400 ppm of \ce{CO2}, and \ce{H2O}, the latter being the main condensible species.
Ozone is not included in our simulations for simplicity, though it may affect our results by modifying the vertical temperature profile in the stratosphere or inhibiting deep convection \citep[see e.g.][]{Gomez-Leal19_climate,Chen19_habitability}.
On the other hand, the radiative influence of ozone is likely to be muted compared to that on Earth, because of the weaker stellar flux in the ozone absorption window \citep[][]{Boutle17_exploring}.
Planetary parameters are the same as in the THAI protocol and are also given here in Table~\ref{tab:planet} for convenience.
The planet is also assumed to be in synchronous rotation, which is justified by the likely time scale of tidal locking compared to the age of its host star \citep{Turbet18_modeling,PierrehumbertHammond19_atmospheric}.
The planet's surface is covered by an immobile slab ocean.
Its bolometric albedo changes depending on the surface temperature as a simple representation of the sea-ice albedo feedback.
The albedo is either 0.06 or 0.25, above or below the freezing point of seawater, respectively.

\subsection{Base and sensitivity experiments}
\label{sec:method_sens}
The \emph{Base} setup is started from an isothermal profile of \SI{300}{\K} and a dry, hydrostatically balanced, atmosphere at rest (first row in Table~\ref{tab:runs}).
Water vapor is then allowed to evaporate from the slab ocean surface and condense into clouds.
Compared to the THAI \textit{Hab~1} UM simulation, here we use a more advanced representation of subgrid cloud variability in the radiation transfer scheme, namely the Monte Carlo Independent Column Approximation (MCICA) while assuming exponential-random overlap \citep{Pincus03_fast,Barker08_mcica}.
To parameterize convection, a mass-flux scheme is used \citep{Gregory90_mass,Walters19_ga7}.
The slab ocean heat capacity is \SI{4e6}{\joule\per\kelvin\per\meter\squared}, corresponding to a depth of \SI{\approx 1}{\meter}, which was also used in previous idealized modeling studies \citep[e.g.][]{Wing18_rcemip,Seeley21_episodic}.
In the \emph{Base} experiment, the surface temperature is allowed to evolve dynamically --- driven by air-sea energy fluxes.

We then run a series of simulations to explore the sensitivity of atmospheric evolution to the initial and boundary conditions, as well as to the choice of the convection parameterization and cloud radiative effects.
The sensitivity experiments are also run for 3000 Earth days, which is sufficient for them to reach a steady state.
We confirmed this by running some of them for 10000 Earth days ($\approx$1639 orbits), during which no further regime transitions happened (not shown).
All simulation setups are summarized in Table~\ref{tab:runs}.

In the first group of experiments, we start the simulation from different temperatures for both the atmosphere and the surface, going from \SI{290}{\K} to \SI{250}{\K}, in increments of \SI{10}{\K}, while holding the rest of the configuration the same as in \emph{Base}.
These experiments are labeled \emph{T0\_290}, \emph{T0\_280}, \emph{T0\_270}, \emph{T0\_260} and \emph{T0\_250}.
In an additional experiment, we test the robustness of the circulation regime by restarting the model in the \emph{Base} configuration from a steady-state snapshot of the DJ regime (labeled \emph{DJ\_start}).

Next, we explore the role of the bottom boundary condition.
We first test the sensitivity of the atmospheric regime to the depth, or equivalently, the heat capacity, of the slab ocean.
In the sensitivity experiments, it is increased to the equivalent depth of \SI{\approx 5}{\meter} and \SI{\approx 10}{\meter} (labeled as \emph{SOD\_5} and \emph{SOD\_10}, respectively)
We then run two experiments with a fixed surface temperature, labeled \emph{FixedSST}.
\footnote{Note that in this configuration the top-of-the-atmosphere (TOA) energy balance fluctuates around a constant non-zero value (which is expected for a fixed temperature setup), implying that the simulation has reached a steady state.}
In \emph{FixedSST\_g}, the surface temperature field is set globally to that obtained in the \emph{T0\_280} experiment (i.e. from the DJ regime).
In \emph{FixedSST\_n}, the DJ-regime surface temperature field is set only to the night hemisphere of the planet, while the day hemisphere surface temperature is fixed to that observed in the SJ regime.
Used extensively in Earth climate modeling \citep[see e.g.][]{Blackburn13_ape,Williamson13_ape}, fixed surface temperature experiments represent a key step in model hierarchy with respect to the lower boundary condition \citep{Maher19_model} and are included in our study to test whether the atmospheric circulation is controlled by the surface thermal forcing on the day or night side of the planet.

With regards to physical parameterizations, we conduct two experiments.
In \emph{Adjust}, we swap the mass-flux convection parameterization for a convection adjustment scheme \citep{Lambert20_continuous}, analogous to the experiment discussed in \citet{Sergeev20_atmospheric}.
Even though convection adjustment schemes are too simplistic to represent the complexity of subgrid-scale convective plumes correctly, they are still often used in modeling planetary atmospheres to understand the key properties of convection \citep[e.g.][]{Lora15_gcm,Labonte20_sensitivity,Turbet21_day-night,Paradise22_exoplasim}.
The \emph{Adjust} experiment thus is designed to test the effect of the representation of convection on the atmospheric circulation.
Similarly, to test the role of cloud radiative feedbacks in the emergence of superrotation, in a separate simulation labeled \emph{CRE\_off} we disable the radiative effect of clouds (both shortwave and longwave).
While somewhat comparable to the benchmark simulations of \citet{Turbet22_thai} with respect to clouds, this experiment still has moisture-related climate processes, such as diabatic heating.

\begin{deluxetable*}{ll}
\tablecaption{Simulation setup. For each sensitivity experiment, only the difference relative to the base experiment is mentioned.}
\label{tab:runs}
\tablewidth{0pt}
\tablehead{
\colhead{Experiment} & \colhead{Description}
}
\startdata
\emph{Base} & Convection scheme: mass-flux\\
            & Cloud radiative effect: ON\\
            & Initial temperature \SI{300}{\K}\\
            & Slab ocean\textsuperscript{\textdagger} depth: \SI{1}{\m}\\
            & Surface temperature: dynamic\\
            & Start: isothermal atmosphere and surface, dry atmosphere, zero wind speed\\
\emph{T0\_250}       & Initial temperature: \SI{250}{\K}\\
\emph{T0\_260}       & Initial temperature: \SI{260}{\K}\\
\emph{T0\_270}       & Initial temperature: \SI{270}{\K}\\
\emph{T0\_280}       & Initial temperature: \SI{280}{\K}\\
\emph{T0\_290}       & Initial temperature: \SI{290}{\K}\\
\emph{DJ\_start} & Start: from a steady-state DJ regime\textsuperscript{\textdagger\textdagger} snapshot\\
\emph{SOD\_5}        & Slab ocean depth: \SI{5}{\m}\\
\emph{SOD\_10}       & Slab ocean depth: \SI{10}{\m}\\
\emph{FixedSST\_g}   & Surface temperature: fixed; DJ regime\textsuperscript{\textdagger\textdagger} distribution globally\\
\emph{FixedSST\_n}   & Surface temperature: fixed; DJ regime\textsuperscript{\textdagger\textdagger} distribution on the night side \\
\emph{Adjust}        & Convection scheme: adjustment\\
\emph{CRE\_off}      & Cloud radiative effect: OFF\\
\enddata
\tablecomments{
\textsuperscript{\textdagger} The slab ocean albedo changes depending on the surface temperature between 0.06 and 0.25, above or below the freezing point of seawater, respectively.}
\textsuperscript{\textdagger\textdagger}``DJ regime'' refers to the double mid-latitude jet circulation pattern described in Sec.~\ref{sec:results_sens}.
\end{deluxetable*}

\subsection{Synthetic spectra}
\label{sec:method_synthobs}
To explore the implications of different climates states for observations, we compute synthetic transmission spectra following the method described in \citet{Lines18_exonephology} and applied for terrestrial planets in \citet{Boutle20_mineral}.
In short, the transmission spectra are calculated using spherical geometry within the 3D GCM framework, using the same radiation scheme (SOCRATES) that the UM uses to simulate the climate.
These calculations use high-resolution (280 bands) spectral files and are performed via a second, ``diagnostic'', call to the radiation scheme thereby not affecting the model evolution.
We do not extend the model top to lower pressures, as has been tested in e.g. \citet{Fauchez22_thai}, as this is not required for a temperate climate with Earth-like temperatures and a low cloud deck \citep{Suissa20_dim,SongYang21_asymmetry}.

\section{Results}
\label{sec:results}
In this section, we present the results of our simulations.
We first show in Sec.~\ref{sec:results_sens} that the SJ and DJ regimes can emerge due to small changes in the model setup.
This makes it important to examine what dynamical mechanisms play a role in the formation and maintenance of the regimes.
Thus, using two illustrative simulations, in Sec.~\ref{sec:results_spinup_dynamics} we focus on the earliest stages of the simulations and show that the evolution of the two regimes is associated with subtle differences in the mean and eddy angular momentum fluxes.
In Sec.~\ref{sec:results_spinup_cold_traps}, we then explain that the large night-side surface temperature difference appears between the regimes due to the difference in water vapor content in the night-side atmosphere.
In Sec.~\ref{sec:results_mean_state}, we explore the climate of these regimes in a steady state and demonstrate that they are well-defined with respect to multiple atmospheric diagnostics, such as surface temperature, wind patters, and cloud distribution.
Finally, in Sec.~\ref{sec:results_synthobs} we discuss the implications for transmission spectra and show that while the circulation regimes have consistent differences in the water absorption bands as well as the continuum level, they are too small to be detected with current technology.

\subsection{Circulation regimes across the model simulations}
\label{sec:results_sens}  
\begin{figure*}
    \centering
    \includegraphics[width=\textwidth]{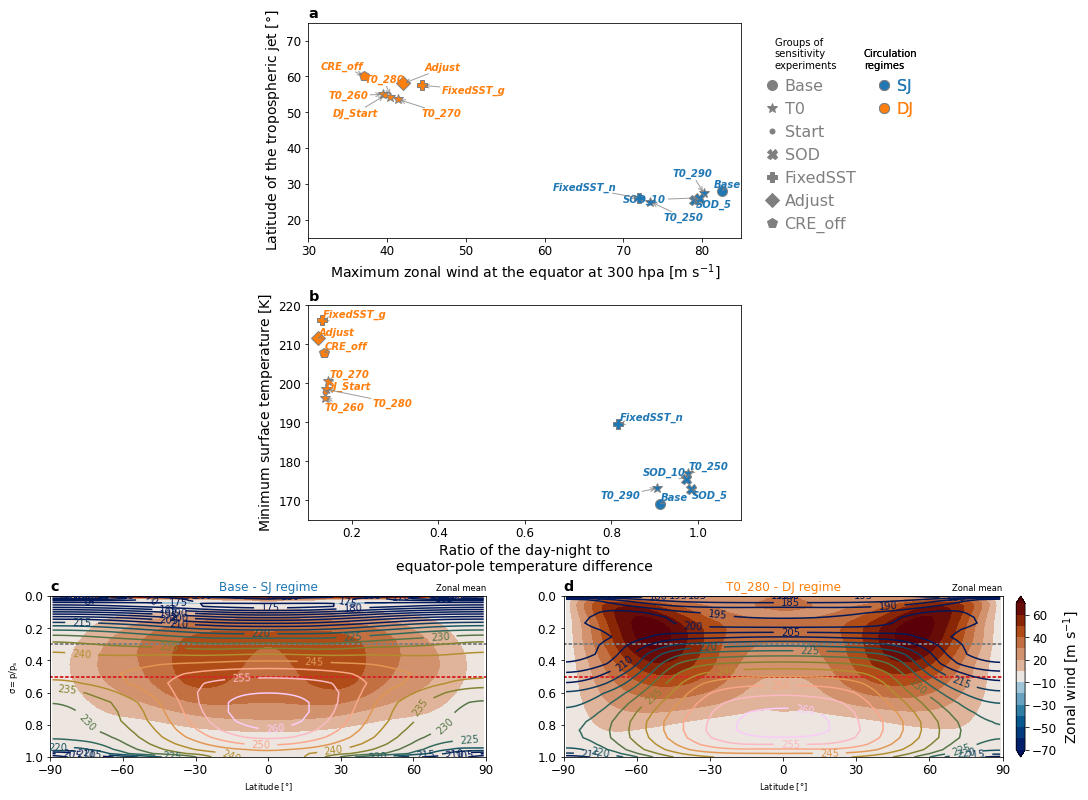}
    \caption{The mean climate diagnostics in all experiments: (a) maximum equatorial zonal wind speed (x-axis, \si{\m\per\s}) and the latitude of the tropospheric jet (y-axis, degrees); (b) the ratio of the day-night to equator-pole temperature difference (x-axis) and the minimum surface temperature (y-axis, \si{\K}).
    Experiments that produce the SJ regime are shown in blue, DJ --- in orange.
    Different marker shapes correspond to different groups of sensitivity experiments.
    Configuration labels are defined in Table~\ref{tab:runs}.
    Also shown is the steady state of the (c) SJ and (d) DJ circulation regimes in the indicative simulations (\emph{Base} and \emph{T0\_280}, respectively).
    Panels c and d show the vertical cross-section of the zonal mean eastward wind (shading, \si{\m\per\s}) and zonal mean air temperature (contours, \si{\K}).
    Horizontal dashed lines in c and d show the corresponding pressure level of the horizontal cross-sections of (red) temperature and (gray) winds and geopotential height shown in Fig.~\ref{fig:mean_circ}.
    ``Single Jet'' (SJ) and ``Double Jet'' (DJ) are short-hand descriptive terms rather than precise descriptions: the equatorial jet in the SJ regime exhibits a split at $\sigma\approx 0.5$, and the DJ regime still has an equatorial superrotation, albeit weaker than that in the SJ regime.
    \label{fig:all_sim}}
\end{figure*}

As described in Sec.~\ref{sec:method_sens}, we include 13 simulations in our study: the \emph{Base} (control) simulation and 12 sensitivity experiments.
In the sensitivity experiments, we change one aspect of the model configuration at a time, keeping the rest of the configuration the same as that in the \emph{Base} setup.
We change the initial conditions (initial temperature), surface boundary conditions (slab ocean depth and temperature), convection parameterization, and cloud radiative effect (CRE).

Fig.~\ref{fig:all_sim} provides a summary of all our experiments in terms of four key diagnostics of the steady-state climate.
These diagnostics are presented in pairs: Fig.~\ref{fig:all_sim}a shows the strength of the eastward wind at the equator and the latitude of the tropospheric jet, while Fig.~\ref{fig:all_sim}b shows the ratio of the day-night to equator-pole temperature difference and the lowest surface temperature.
It is apparent in both panels that the experiments form two distinct clusters, and there is practically no spectrum between the climate regimes.
Note that the clusters of experiments are the same for all four metrics.

The SJ regime has a higher zonal wind in the equatorial upper troposphere, with its maximum reaching values \SI{\approx 80}{\m\per\s} (x-axis in Fig.~\ref{fig:all_sim}a).
The low values of the jet latitude in this regime demonstrate that the zonal wind maximum is in the tropics (y-axis in Fig.~\ref{fig:all_sim}a).
The DJ regime, on the other hand, has substantially lower zonal wind at the equator --- at about \SI{\approx 40}{\m\per\s}.
However, the DJ regime still maintains an equatorial superrotation, albeit a weaker one compared to that in the SJ regime (Fig.~\ref{fig:all_sim}c,d).
The maximum of the zonal wind speed in the DJ regime is at \ang{\approx 60} latitude, demonstrating that the dominant tropospheric jets are extratropical.

The thermal structure of the SJ regime is such that the temperature difference between the day and night side of the planet is largely equal to the equator-to-pole temperature gradient (x-axis in Fig.~\ref{fig:all_sim}b).
This is mostly due to a colder night side, illustrated by the relatively low surface temperatures in night-side ``cold traps'' in this regime (\SI{<180}{\K}, see the y-axis in Fig.~\ref{fig:all_sim}b).
Note that the only outlier is the \emph{FixedSST\_n} simulation because its night-side temperature is fixed to that of the DJ regime, i.e. a higher value.
Indeed, the DJ regime has consistently higher night-side surface temperatures --- between 220 and \SI{230}{\K}, than that found in SJ simulations.
Consequently, the meridional temperature gradient between the equator and poles is more than 5 times larger than the day-night contrast (x-axis in Fig.~\ref{fig:all_sim}b).
In other words, the DJ thermal structure is more zonally symmetric than that of the SJ regime.
The details of the thermodynamic and circulation patterns of both regimes are discussed in more detail in Sec.~\ref{sec:results_mean_state}.

Fig.~\ref{fig:all_sim} also reveals which of the two climate regime the simulation are sensitive to the following factors: initial conditions overall (\emph{DJ\_start}) and temperature in particular (\emph{T0} group), fixed surface temperature distribution (\emph{FixedSST\_g}), the choice of the convection scheme (\emph{Adjust}), and the inclusion or omission of the radiative impact of clouds (\emph{CRE\_off}).
When one of these aspects of the model setup is changed, the resulting simulation is in the DJ regime (opposite to that in \emph{Base}).
On the other hand, the circulation regime is insensitive to changing the slab ocean depth (experiments \emph{SOD\_5} and \emph{SOD\_10}) or the night-side surface temperature (\emph{FixedSST\_n}).

An important sensitivity simulation is \emph{DJ\_start}, i.e. the simulation started not from an isothermal profile and an atmosphere at rest, but from a previously developed DJ regime.
This steady state snapshot is taken from the \emph{Adjust} simulation and includes all the prognostic model fields \citep{Walters19_ga7}, such as the wind components, atmospheric pressure, temperature, water vapor and cloud content.
As Fig.~\ref{fig:all_sim} demonstrates, such initial conditions appear to determine the end state: the already established DJ regime does not spontaneously transition to the SJ regime, even if the convection scheme is not that used in the \emph{Adjust} case.
This hints at the fact that the DJ regime is more robust than its SJ counterpart, but further conclusions require a separate study.

The simulations started from a different initial temperature $T_0$ provide the most interesting outcome of our model sensitivity study, because moderate initial temperatures (260, 270, and \SI{280}{\K}) result in the DJ regime, while the extremes give the SJ regime (e.g. \emph{T0\_290} with $T_0=\SI{290}{\K}$ and \emph{Base} with $T_0=\SI{300}{\K}$).
It is also the most surprising result, because one would not expect large sensitivity to $T_0$ set within reasonable limits: our simulations do not include a dynamical ocean or sophisticated sea ice schemes \citep[see e.g.][]{DelGenio19_habitable,Yang20_transition,Olson22_effect}, and the slab ocean provides an infinite source of moisture (which would be important only on large time scales).
The \SI{300}{\K} isothermal initial state specified by the THAI protocol, which our control setup inherits, was chosen primarily for its simplicity \citep{Fauchez20_thai_protocol}, but no systematic investigation of the model sensitivity was performed.

In the remaining sections, we deliver a detailed comparison of the SJ and DJ regimes, focusing on the \emph{Base} and \emph{T0\_280} simulations, respectively.
We show that even a \SI{20}{\K} initial temperature difference can lead to a regime bistability within the first tens of days of model evolution.
It happens due to the different amount of water vapor lifted in the atmosphere by convection in the substellar region, which results in a different radiative forcing of the atmosphere and further consequences for superrotation and the overall climate.


\subsection{Emergence of the circulation regimes}
\label{sec:results_spinup_dynamics}
\begin{figure*}
    \centering
    \includegraphics[width=\textwidth]{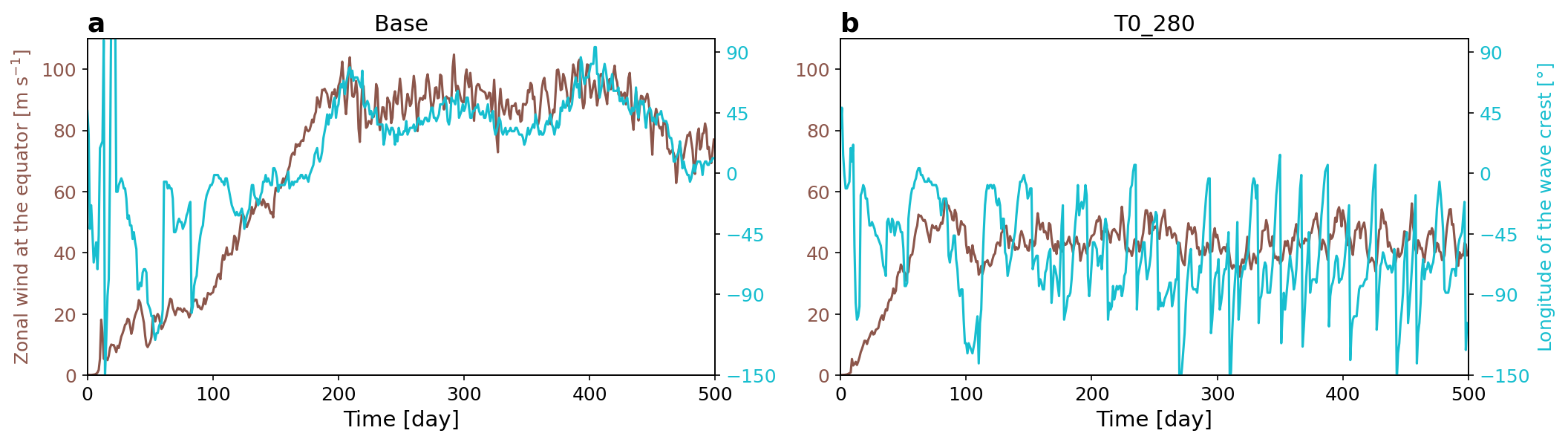}
    \caption{Spin-up diagnostics for the (a) \emph{Base} (b) \emph{T0\_280} simulations: (brown) maximum zonal wind at the equator at \SI{300}{\hecto\pascal} and (cyan) the longitude of stationary Rossby wave crest.
    For reference, the wave crest longitude is also shown in Fig.~\ref{fig:mean_circ}c,d.
    An animation of the stationary wave pattern and zonal mean atmospheric structure during the first 500 days of the simulations is provided as Supplemental Video 1 (see also Fig.~\ref{fig:spinup_ghgt_anom_wind_temp}).
    \label{fig:crest_lon_eq_jet}}
\end{figure*}

While many previous studies discuss the maintenance of superrotation on tidally locked exoplanets in a steady state regime \citep[e.g.][]{ShowmanPolvani11_equatorial,Tsai14_three-dimensional,Carone15_connecting,KomacekShowman16_atmospheric,Noda17_circulation}, its initial acceleration received much less attention, especially in a full-complexity atmospheric GCM.
For hot Jupiter atmospheres, it was explored in 3D GCM simulations by e.g. \citet{Liu13_atmospheric} and \citet{Debras20_acceleration}.
\citet{WangYang20_phase} also briefly discussed it in the context of a wave-jet resonance on a hypothetical tidally locked terrestrial planet.
Building on these studies, we scrutinize the initial phase of the two regimes and track the development of the wave-mean-flow interaction.
The regime evolution described here is unlikely to happen in a real atmosphere, because no atmosphere develops from quiescent isothermal conditions.
However, a small change in forcing (for example due to a stellar flare) on TRAPPIST-1e or a similar exoplanet, whose atmosphere resides on the edge of different regimes, may result in an abrupt change in circulation with consequences for the global climate \citep[e.g.][]{SuarezDuffy92_terrestrial,Caballero10_spontaneous,Arnold12_abrupt,Noda17_circulation}.
It is also crucial to understand how the two different circulation regimes develop in a 3D GCM in order to be confident in the robustness of GCM simulations of an exoplanetary climate.
This will allow for more informed decisions in setting up future single-model studies and GCM intercomparisons \citep{Fauchez21_workshop}.

The evolution of the flow is summarized in Fig.~\ref{fig:crest_lon_eq_jet} by the daily-mean time series of the zonal wind at \SI{300}{\hecto\pascal} along with the phase of the stationary Rossby wave.
The latter is diagnosed by the longitude of the maximum of \SI{300}{\hecto\pascal} eddy geopotential height, i.e. the deviation from the zonal mean of the \SI{300}{\hecto\pascal} isobaric surface height.
It takes between 100 to 250 Earth days for equatorial superrotation to settle into either the SJ or DJ regime in the \emph{Base} and \emph{T0\_280} case, respectively.
Note in the sensitivity experiments with a deeper slab ocean (i.e. with higher heat capacity), the flow evolution takes longer to stabilize (not shown), but the final state does not differ from the control (Fig.~\ref{fig:all_sim}).

The wind speed time series in Fig.~\ref{fig:crest_lon_eq_jet} show that within the first $\approx$80 days, the equatorial superrotation developed in the \emph{Base} setup is weaker than that in the \emph{T0\_280} case (see also Supplemental Video 1 and Fig.~\ref{fig:spinup_ghgt_anom_wind_temp}).
During this first acceleration stage, the planetary-scale wave pattern also develops quicker in the \emph{T0\_280} case and is able to transport eastward momentum to the equator, accelerating the jet to higher velocity relative to that in \emph{Base}.
After approximately day 80, in the \emph{T0\_280} simulation the broad equatorial superrotating flow splits into two separate jet cores, which migrate to mid-latitudes and within a few further days reach their steady-state structure --- the DJ regime.
Accordingly, the eastward momentum supplied to these jets is being taken from the equatorial region, causing the zonal wind at the equator to slow down to \SI{\approx 45}{\m\per\s} (Fig.~\ref{fig:crest_lon_eq_jet}b).
The zonal flow thus fails to achieve resonance with the stationary Rossby wave, whose crest keeps oscillating near the western terminator \citep[see e.g.][]{PierrehumbertHammond19_atmospheric,WangYang20_phase}.
Meanwhile, in the \emph{Base} experiment, the equatorial superrotation continues to develop more gradually and reaches its steady-state maximum by approximately day 200.
A wave-jet resonance develops, which is seen in the acceleration of the equatorial jet to \SI{\approx 90}{\m\per\s} and an eastward shift of the planetary wave, whose crest settles at \ang{\approx 45}E (Fig.~\ref{fig:crest_lon_eq_jet}a, see also Supplemental Video 1 and Fig.~\ref{fig:spinup_ghgt_anom_wind_temp}).

The period needed to reach the steady state is comparable to those found in the idealized experiments of \citet{Noda17_circulation} and \citet{HammondPierrehumbert18_wave-mean}.
The manifestation of the wave-jet resonance in the \emph{Base} case is also similar to that shown in \citet{WangYang20_phase}, though happens over a longer period of time, likely because of the uniform initial conditions in our setup.
Notably, the wind speed and wave crest longitude exhibit oscillations around the steady state.
This time variability is more prominent in the \emph{T0\_280} case, because it is associated with a larger role of transient baroclinic eddies, as we discuss further below.

\begin{figure*}
    \centering
    \includegraphics[width=0.8\textwidth]{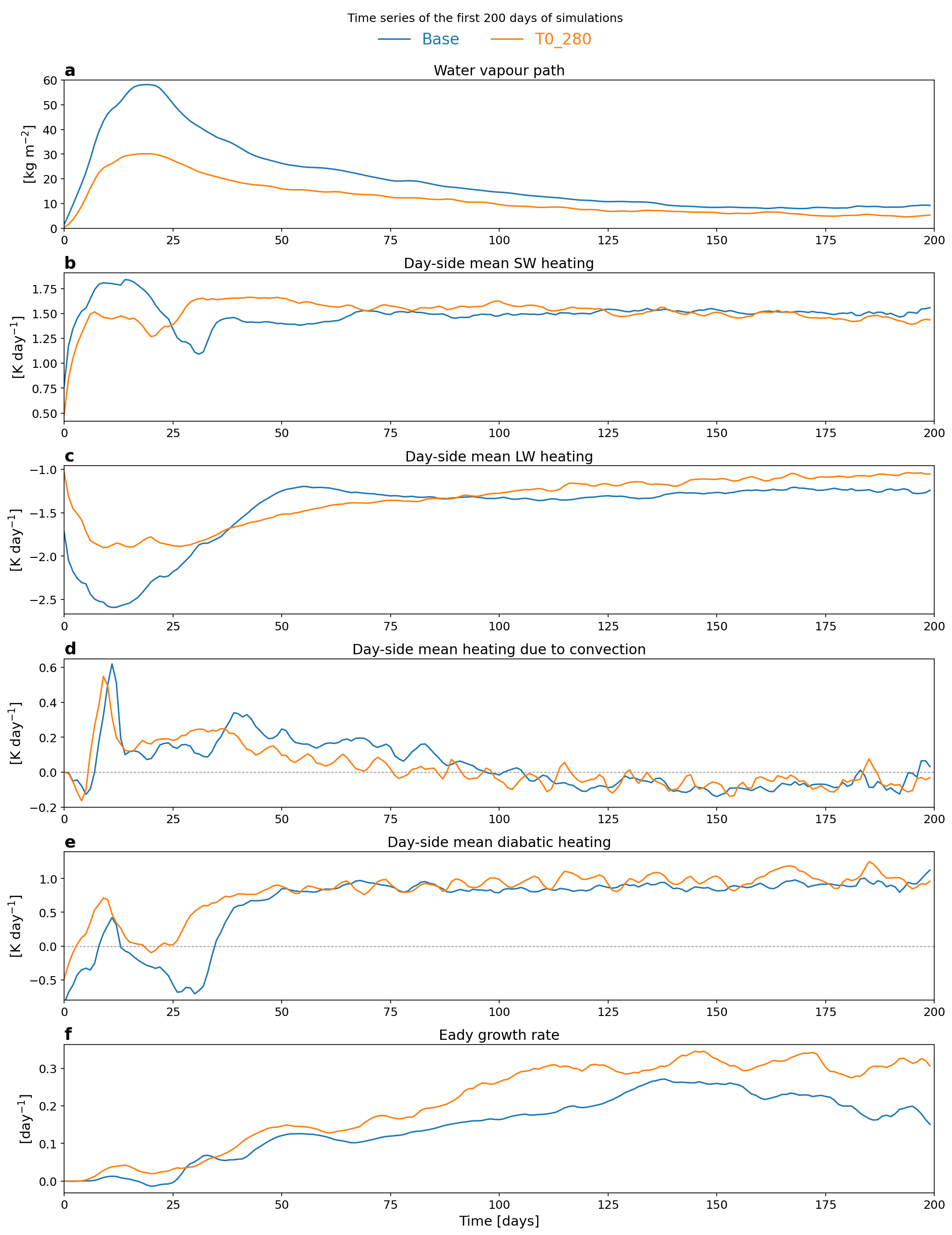}
    \caption{Spin-up diagnostics for the first 200 days of the (blue lines) \emph{Base} and (orange lines) \emph{T0\_280} simulations: (a) water vapor path, (b) shortwave radiative heating rate, (c) longwave radiative heating rate, (d) convective heating rate, (e) diabatic (radiative plus latent) heating rate, (f) the Eady growth rate.
    The water vapor path is averaged over the day side, integrated vertically and has units of \si{\kg\per\m\squared}.
    The heating rates are averaged spatially over the day side and vertically over the troposphere and have units of \si{\K\per\day}.
    The Eady growth rate is averaged within \SIrange{\approx 850}{500}{\hecto\pascal} in mid-latitudes (\ang{30}--\ang{80}) and has units of \si{\per\day}.
    \label{fig:var_tseries}}
\end{figure*}

The regime bifurcation within the first tens of days of the two simulations can be explained using the diagnostics in Fig.~\ref{fig:var_tseries}.
In the \emph{Base} case, the high initial temperature of the surface (\SI{300}{\K}), further increased due to stellar irradiation initially unimpeded by clouds, leads to extremely strong surface latent heat flux via evaporation.\footnote{Note that the initial temperature spike beyond \SI{300}{\K} may not be crucial, because the simulation with a fixed surface temperature, \emph{FixedSST\_g}, settles into the SJ regime (Fig.~\ref{fig:all_sim}).}
The strong latent heat flux induces vigorous convection, manifesting at day 10 as a spike of convective heating of up to \SI{0.6}{\K\per\day}, evident in Fig.~\ref{fig:var_tseries}d (blue curve).
Convective plumes at the substellar region lift significant volumes of water vapor into the atmosphere.
The atmosphere's high initial temperature (also \SI{300}{\K}) allows it to hold a large portion of that moisture before it condenses, as described by the Clausius-Clapeyron equation.
The \emph{Base} case thus experiences a marked increase in the total column water vapor (water vapor path) --- up to \SI{60}{\kg\per\m\squared} --- within the first 20 days (Fig.~\ref{fig:var_tseries}a).

Moistening of the substellar atmosphere produces an increase in shortwave absorption to more than \SI{1.8}{\K\per\day} (Fig.~\ref{fig:var_tseries}b).
However, the water vapor also efficiently radiates energy to space, causing the longwave cooling rate to reach \SI{\approx 2.5}{\K\per\day} (Fig.~\ref{fig:var_tseries}c).
The result is the net cooling of the atmosphere by radiation.
The day-side radiative cooling is offset by heating due to deep convection (Fig.~\ref{fig:var_tseries}d), turbulent fluxes in the boundary layer, and condensation of the water vapor.
The contribution from the latter two processes is smaller relative to convection and thus is not shown.
Consequently, the total forcing of the day side atmosphere is weakly negative in the \emph{Base} case until approximately day 40, after which it increases to the steady-state value of \SI{\approx 1}{\K\per\day} (Fig.~\ref{fig:var_tseries}e).

The \emph{T0\_280} simulation, on the other hand, starts from a profile \SI{20}{\K} colder than that in the \emph{Base} case.
As a result, the surface evaporation and deep atmospheric convection is slightly weaker (orange curve in Fig.~\ref{fig:var_tseries}d).
Furthermore, the saturation water vapor pressure is also lower due to the atmosphere being colder, and the resulting increase in the total column water vapor is about half as much as in the \emph{Base} simulation (Fig.~\ref{fig:var_tseries}a).
While the difference in the shortwave heating between the two experiments is small, the difference in the longwave cooling is larger, with the \emph{T0\_280} atmosphere losing \SI{< 2}{\K\per\day}.
As far as the total latent heating is concerned, it is overall similar in both cases, as exemplified by the convective heating rate in Fig.~\ref{fig:var_tseries}d.
It is also smaller than the radiative heating rates, in agreement with \citet{Boutle17_exploring}.
The overall effect is thus mainly due to the differences in the radiative heating rates, making the net diabatic forcing in the \emph{T0\_280} simulation stronger than that in the control one (Fig.~\ref{fig:var_tseries}e).

This is likely the key difference in the initial stages of the two simulations that places them on different branches of regime evolution.
Namely, the overall weaker forcing in the \emph{Base} simulation, relative to that in \emph{T0\_280}, produces a slower development of the stationary wave and a more gradual acceleration of the equatorial eastward jet (Fig.~\ref{fig:crest_lon_eq_jet}a).
On the contrary, the stronger forcing in the \emph{T0\_280} case establishes the wave pattern and accelerates the equatorial jet (initially) more rapidly (Fig.~\ref{fig:crest_lon_eq_jet}b).
These results agree with the earlier studies based on shallow water models as well as idealized GCMs.
For example, \citet{HammondPierrehumbert18_wave-mean} show the GCM output for a dry tidally locked terrestrial planet with a 5 day rotation period (close to that of TRAPPIST-1e, see Table~\ref{tab:planet}).
The authors demonstrate that for a fixed planetary rotation rate, different circulation regimes emerge depending on the strength of stellar forcing \citep[see also][]{HammondPierrehumbert18_wave-mean}.
Qualitatively, their regime at the highest instellation is similar to that emerging in the \emph{T0\_280} simulation, the defining feature of which is a single broad equatorial eastward jet and a high-amplitude planetary-scale wave.
At the lowest instellation, the authors obtain a regime similar to that in the initial phase of the \emph{Base} simulation with a weaker equatorial superrotation.
Note the change of stellar forcing between the regimes in their study is substantially larger than the changes in forcing in the first days of our simulations.
This is merely a qualitative comparison, however, because \citet{HammondPierrehumbert18_wave-mean} analyze the steady-state circulation, not the acceleration phase.
As discussed below, even though the \emph{Base} simulation starts with a weaker equatorial superrotation, it ends up with a stronger superrotation; whilst the \emph{T0\_280} simulation starts with a stronger superrotation, but ends up with a weaker one.

After this initial development phase ($\approx$80 days), the SJ-like circulation pattern in the \emph{T0\_280} case transforms into the DJ circulation pattern by developing a pair of eastward jets at mid-latitudes.
This corresponds to the decrease in the stationary wave amplitude and the deceleration of the equatorial jet (Fig.~\ref{fig:crest_lon_eq_jet}b).
In other words, the \emph{T0\_280} simulation fails to achieve a wave-jet resonance.
Instead, the \emph{T0\_280} case is characterized by an increase in baroclinicity manifested as baroclinic waves traveling in the zonal direction at high latitudes (see Supplemental Video 1).
The increasing role of baroclinic instability, especially after the first 80 days, is demonstrated by the time series of the Eady growth rate which is calculated following \citet{Vallis17_aofd}:
\begin{equation}
    \sigma_E = 0.31|f|\frac{|\partial u/\partial z|}{N},
\end{equation}
where $f=2\Omega sin\phi$ is the Coriolis parameter ($\Omega$ is the planetary rotation rate, $\phi$ is the latitude), $\partial u/\partial z$ is the derivative of the zonal wind velocity with height, $N=\sqrt{g/\theta\partial\theta/\partial z}$ is the Brunt-V\"{a}is\"{a}l\"{a} frequency ($g$ is the acceleration due to gravity, $\theta$ is the potential temperature).
Fig.~\ref{fig:var_tseries}f shows that $\sigma_E$ is consistently higher for the \emph{T0\_280} than for the \emph{Base} simulation (orange and blue curves, respectively).
This indicates that the \emph{T0\_280} case develops conditions more favorable for the baroclinic instability, mostly via the increase of the mean horizontal temperature gradient, which via the thermal wind equation is proportional to $\partial u/\partial z$.
The emergence of baroclinic jets at mid-latitudes marks the mature stage of the DJ regime.

In the \emph{Base} experiment, $\sigma_E$ is substantially smaller, indicating a weaker role of baroclinic instability (Fig.~\ref{fig:var_tseries}f).
The SJ regime reaches its equilibrium and does not develop a strong equator-pole temperature gradient (Fig.~\ref{fig:all_sim}).
Broadly the same chain of events leading to one regime or another is identified across the other sensitivity experiments in our study.
One interesting example is \emph{T0\_250}, which eventually develops an SJ regime, despite its colder initial conditions.
Despite the colder start, which favors stronger initial forcing and thus the evolution similar to that in the \emph{T0\_280} case, the \emph{T0\_250} simulation does not develop strong baroclinicity in mid-latitudes and thus eventually transitions back to the SJ regime (not shown).

From the dynamical perspective, the zonal flow acceleration can be analyzed using the zonal component of the axial angular momentum budget.
Hereafter simply referred to as angular momentum, it is defined per unit mass as $m = (u + \Omega r\cos\phi)r\cos\phi$ where $u$ is the zonal wind speed, $\Omega$ is the rotation rate, $r$ is the planetary radius, and $\phi$ is latitude.
The time and zonal mean budget of $m$, without a shallow atmosphere approximation, can be expressed as
\begin{align}
    \frac{\Delta[\rho m]}{\Delta T}&=
    \underbrace{ - \frac{[\overline{V}]}{r}\frac{\partial[\overline{m}]}{\partial\phi} }_\text{Term MH}
    \underbrace{ - [\overline{W}]\frac{\partial[\overline{m}]}{\partial r} }_\text{Term MV}
    \nonumber\\
    &\phantom{=\ \,}
    \underbrace{ -\frac{1}{r\cos\phi}\frac{\partial}{\partial\phi}([\overline{V}^{\ast}\overline{m}^{\ast}]\cos\phi) }_\text{Term SH}
    \underbrace{ - \frac{1}{r^{2}}\frac{\partial}{\partial r}([\overline{W}^{\ast}\overline{m}^{\ast}]r^{2}) }_\text{Term SV}
    \nonumber\\ 
    &\phantom{=\ \,} 
    \underbrace{ -\frac{1}{r\cos\phi}\frac{\partial}{\partial\phi}([\overline{V^{\prime}m^{\prime}}]\cos\phi) }_\text{Term TH}
    \underbrace{ - \frac{1}{r^{2}}\frac{\partial}{\partial r}([\overline{W^{\prime}m^{\prime}}]r^{2}) }_\text{Term TV}
    \nonumber\\
    &\phantom{=\ \,} - r\cos\phi[\overline{\rho G_\lambda}],
    \label{eq:aam_main}
\end{align}
where square brackets denote zonal mean and overbars denote time mean, while asterisks and primes denote the deviations from the zonal and time mean, respectively.
The term on the left-hand side is $\Delta[\rho m] = ([\rho m]_{t=\Delta T} - [\rho m]_{t=0})$ divided by $\Delta T$, which is the total change in $m$ over the time period $\Delta T$.
The rest of the notations are as follows: $\rho$ is density, $V=\rho v$ and $W=\rho w$, where $v$ and $w$ are the meridional and vertical wind speeds, respectively; $G_\lambda$ represents friction and dissipation forces.
The derivation of Eq.~\eqref{eq:aam_main} is given in Appendix~\ref{app:aam}.
\eqref{eq:aam_main} states that the change in mean angular momentum can be due to three transport components, each of which can be split into horizontal and vertical parts (H and V, respectively).
The first two terms on the right hand side (MH and MV) represent the advection of mean $m$ by the mean flow, the third and the fourth terms (SH and SV) represent the transport by stationary eddies, while the fifth and the sixth terms (TH and TV) represent the transport by transient eddies.
Note that the mean terms (MH and MV) are written in the advective form.
Eq.~\eqref{eq:aam_main} has a form similar to that for the zonal wind $u$ used in many previous studies \citep[e.g.][]{Kraucunas05_equatorial,Hammond20_equatorial,ZengYang21_oceanic}, but has a more concise form by inherently incorporating the Coriolis force terms within $m$.

\begin{figure*}
    \centering
    \includegraphics[width=0.8\textwidth]{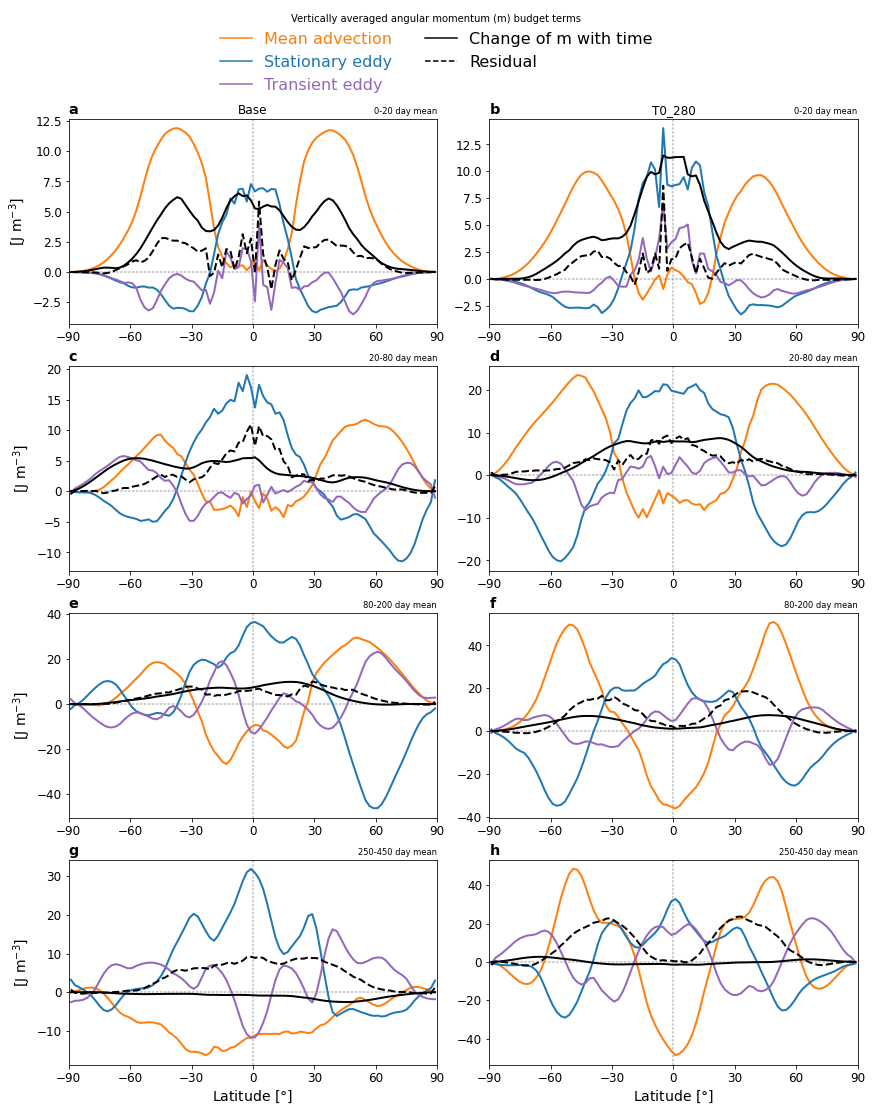}
    \caption{Meridional profiles of the angular momentum budget terms (\si{\joule\per\m\cubed}) in Eq.~\ref{eq:aam_main} during the spin-up phase of the (left) \emph{Base} and (right) \emph{T0\_280} simulations: (orange) mean advection terms, (blue) stationary eddy terms, (purple) transient eddy terms.
    The dashed black line shows the residual.
    The terms are averaged within the troposphere (\SIrange{\approx 1}{20}{\km}) and over the periods of (a, b) 0--20 days, (c, d) 20--80 days, (e, f) 80--200 days, and (g, h) 250--450 days.
    Note the jagged lines in the two top panels are due to a very short period of averaging (20 days).}
    \label{fig:ang_mom_yprof}
\end{figure*}

Fig.~\ref{fig:ang_mom_yprof} shows the meridional profiles of the angular momentum budget terms calculated according to Eq.~\ref{eq:aam_main} over four periods of the flow evolution.
During the first stage (0--20~days of the simulation), the dominant terms are the mean advection terms and and are maximized in extratropical latitudes (Fig.~\ref{fig:ang_mom_yprof}a,b).
This is mostly due to the horizontal Coriolis acceleration, which is positive in the mid-troposphere due to the strong meridional divergence of the flow. 
This term is roughly the same in both \emph{Base} and \emph{T0\_280} simulations.
The eddy angular momentum transport, however, is notably higher in the \emph{T0\_280} case (Fig.~\ref{fig:ang_mom_yprof}b), corresponding to a stronger acceleration of the equatorial eastward jet (Fig.~\ref{fig:crest_lon_eq_jet}b).
Most of the eddy transport is due to the stationary terms, which transport momentum horizontally from the tropics and mid-latitudes toward the low latitudes and upwards to the upper troposphere at the equator (not shown).
A weak stationary eddy contribution in the \emph{Base} case and a strong one in the \emph{T0\_280} case is in agreement with the initial forcing being likewise weaker and stronger in these simulations.
One can notice that the residual is large in Fig.~\ref{fig:ang_mom_yprof}a, b.
This is likely due to the fact that the simulations are started from rest and the mean-eddy separation is not clear during the earliest stages of the model spin-up.
Another possible source of error is sampling rate: we use daily mean output, which likely leads to an underestimation of the eddy terms.

In the days 20--80 of the simulations, the day-side mean diabatic heating is still stronger in the \emph{T0\_280} case than that in the \emph{Base} case, explaining the slightly stronger stationary eddy transport to the equator (Fig.~\ref{fig:ang_mom_yprof}c,d).
Meanwhile, the mean advection terms increase in magnitude compared to those in the \emph{Base} case and form prominent peaks at mid-latitudes (Fig.~\ref{fig:ang_mom_yprof}d).
The role of transients in this period is small relative to the mean and stationary contributions.

During the next period (80--200~days), while the day-side forcing reaches a steady-state (Fig.~\ref{fig:var_tseries}e), the \emph{Base} case has a weak and meridionally asymmetric acceleration of the zonal flow, indicating that the circulation structure is not yet stable (Fig.~\ref{fig:ang_mom_yprof}e, see also Supplemental Video 1).
The $m$ budget in the \emph{T0\_280} case, on the other hand, experiences a doubling of the magnitude of the mean transport terms and a decrease of the stationary term magnitude at the equator.
This reflects the fact that the balance is tipped in favor of the mean transport of the angular momentum to high latitudes instead of its transport by eddies to the equator (Fig.~\ref{fig:ang_mom_yprof}e).
Accordingly, the equatorial jet decelerates, while the pair of mid-latitude jets accelerates.

By the end of this period (at $\approx$200~days), the equatorial jet in the \emph{Base} experiment increases to its steady state level (Fig.~\ref{fig:crest_lon_eq_jet}a), locking in a resonance with the stationary wave pattern.
This happens as the eastward flow approaches the phase velocity of the wavenumber 1 Rossby wave mode (with the opposite sign), which in our simulations is close to \SI{80}{\m\per\s} estimated according to \citet{WangYang20_phase}.
As this flow speed approaches this threshold, the free Rossby mode becomes stationary relative to the heating in the substellar region and amplifies in magnitude.
Dampened by friction, the wave amplification reaches its maximum when the zonal wind is equal to the Rossby wave speed, which can be thought of as a resonance \citep{Arnold12_abrupt}.
Note that the Kelvin wave speed is much higher and directed opposite to the mean flow in our simulations, so a resonance with the Kelvin wave is not relevant \citep{WangYang20_phase}.

Fig.~\ref{fig:ang_mom_yprof}g and h show contributions to the $m$ budget from each of the terms in Eq.~\ref{eq:aam_main} at equilibrium (beyond 250~days).
The total change of the angular momentum is close to zero, as indicated by the solid black curves, and the mean and eddy terms largely cancel each other out.
In the \emph{T0\_280} case, the shape and the magnitude of the $m$ budget terms remains similar to those in the previous time period, only intensifying the angular momentum transport from the equator by the mean circulation (Fig.~\ref{fig:ang_mom_yprof}h).
Meanwhile, the same mean transport term in the \emph{Base} case decreases substantially, approaching zero at mid-latitudes.
This term's negative values at low latitudes are balanced by the positive stationary eddy term.
Evidently, the stationary term continues to transport angular momentum equatorward, drawing it from high latitudes where it is replenished by the transient eddy term (Fig.~\ref{fig:ang_mom_yprof}g).
The horizontal stationary eddy flux of angular momentum converges in the upper troposphere and diverges in the mid-troposphere, resulting in positive and negative contributions to the momentum budget, respectively (not shown).
The redistribution of angular momentum from the upper layers to the deep layers is performed by the vertical component of the stationary eddy term.
Supporting these findings, the same pattern of eddy acceleration was associated with equatorial superrotation in previous studies of tidally locked planets, assuming various atmospheric conditions and various degree of model complexity \citep[e.g.][]{Tsai14_three-dimensional,Showman15_three-dimensional,HammondPierrehumbert18_wave-mean,Debras20_acceleration,Hammond20_equatorial}.

To sum up, the initial evolution of the SJ and DJ regimes is not monotonic and is driven by a combination of mean overturning circulation, concomitant with higher baroclinicity at mid-latitudes, and zonally asymmetric planetary-scale forcing (with a maximum in the substellar region) due to the planet's synchronous rotation.
Both processes compete in our simulations, and their relative strength during the first 100--200 days determines the trajectory leading to one distinct regime or another.
In the \emph{Base} case, which eventually settles on the SJ regime, the acceleration of the equatorial superrotation is slow and steady, because of a weaker day side radiative forcing, which in turn is damped by the relatively strong longwave cooling due to the high concentration of water vapor.
Nevertheless, the SJ regime is eventually realized, as the equatorial eastward jet reaches the Rossby wave speed, indicating a wave-jet resonance.
The resonance-amplified Rossby wave maintains an excess of angular momentum at the equator, i.e. superrotation.
The \emph{T0\_280} case, while initially developing a strong equatorial jet reminiscent of the SJ regime, experiences a transition to the DJ regime after about 80 days.
The initial equatorial jet acceleration may be attributed to the day side forcing being stronger than that in the \emph{Base} case, while the subsequent transition to the DJ regime is driven by enhanced poleward fluxes of angular momentum due to the mean flow.
It is difficult to pinpoint the root cause for the regime bifurcation in a complex GCM such as the UM.
To explain why a more gradual jet acceleration leads to a SJ regime consistent with a Rossby wave resonance (in the \emph{Base} case), while a more rapid jet acceleration leads to a DJ regime (in the \emph{T0\_280} case), one would likely need a more idealized GCM with an option to prescribe forcing and emulate the regime evolution shown here in a more controlled environment.

The difference in the stationary wave pattern between the regimes is also associated with the position of the night-side cyclonic gyres with cold surface temperatures underneath.
The time evolution of the night-side temperature minima is discussed in detail in the next section (Sec.~\ref{sec:results_spinup_cold_traps}).
Further description of the steady state climate in the SJ and DJ regimes, is given in Sec.~\ref{sec:results_mean_state}, and their imprint in the transmission spectrum --- in Sec.~\ref{sec:results_synthobs}.

\subsection{The night side surface temperature evolution in the two regimes}
\label{sec:results_spinup_cold_traps}
\begin{figure*}
\includegraphics[width=\textwidth]{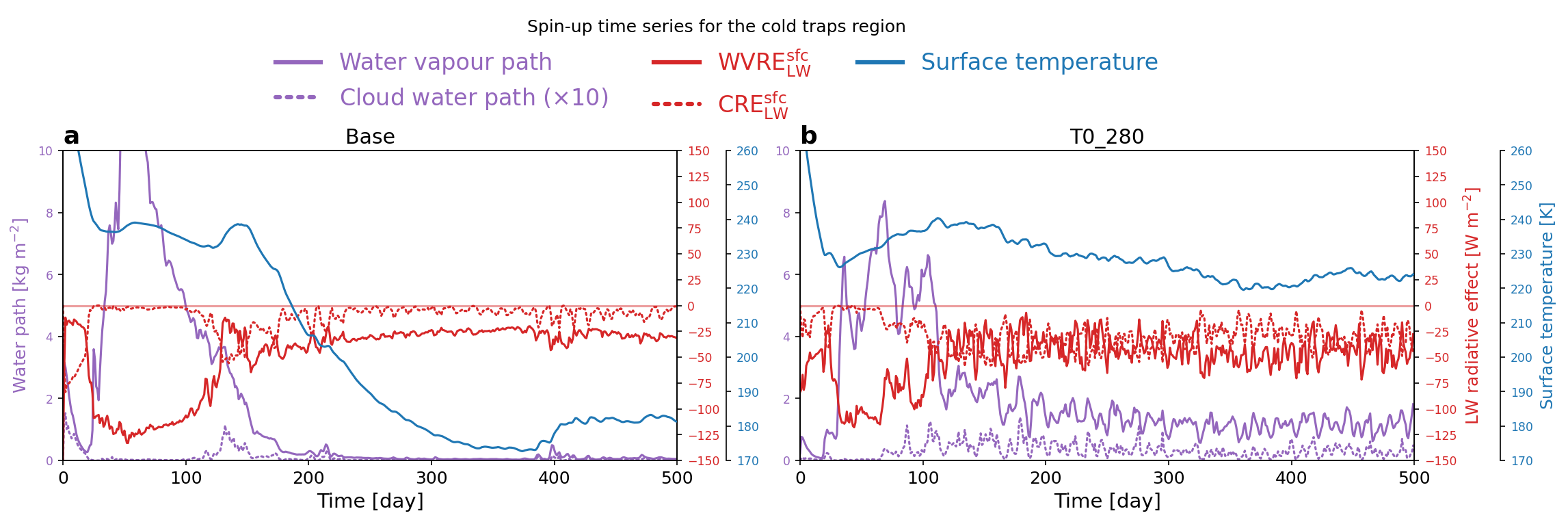}
\caption{Time series of diagnostics for the night-side cold traps, defined as the region bounded by \ang{45} and \ang{55} in the latitude and \ang{160}--\ang{140}W in the longitude. The panels for the (left) SJ and (right) DJ regime show: (purple, solid) water vapor path in \si{\kg\per\m\squared}, (purple, dashed) cloud water path in \SI{10}{\kg\per\m\squared}, (red, solid) water vapor radiative effect WVRE\textsubscript{LW}\textsuperscript{sfc} in \si{\watt\per\m\squared}, (red, dashed) cloud radiative effect CRE\textsubscript{LW}\textsuperscript{sfc} in \si{\watt\per\m\squared}, and (blue) surface temperature in \si{\K}. The WVRE\textsubscript{LW}\textsuperscript{sfc} is defined as the difference between the radiative fluxes at the surface calculated with and without the water vapor opacity. Likewise, the CRE\textsubscript{LW}\textsuperscript{sfc} is defined as the difference between the ``clear-sky'' and ``cloudy'' radiative fluxes at the surface.\label{fig:cold_traps_wvre_cre}}
\end{figure*}
\begin{figure*}
\includegraphics[width=\textwidth]{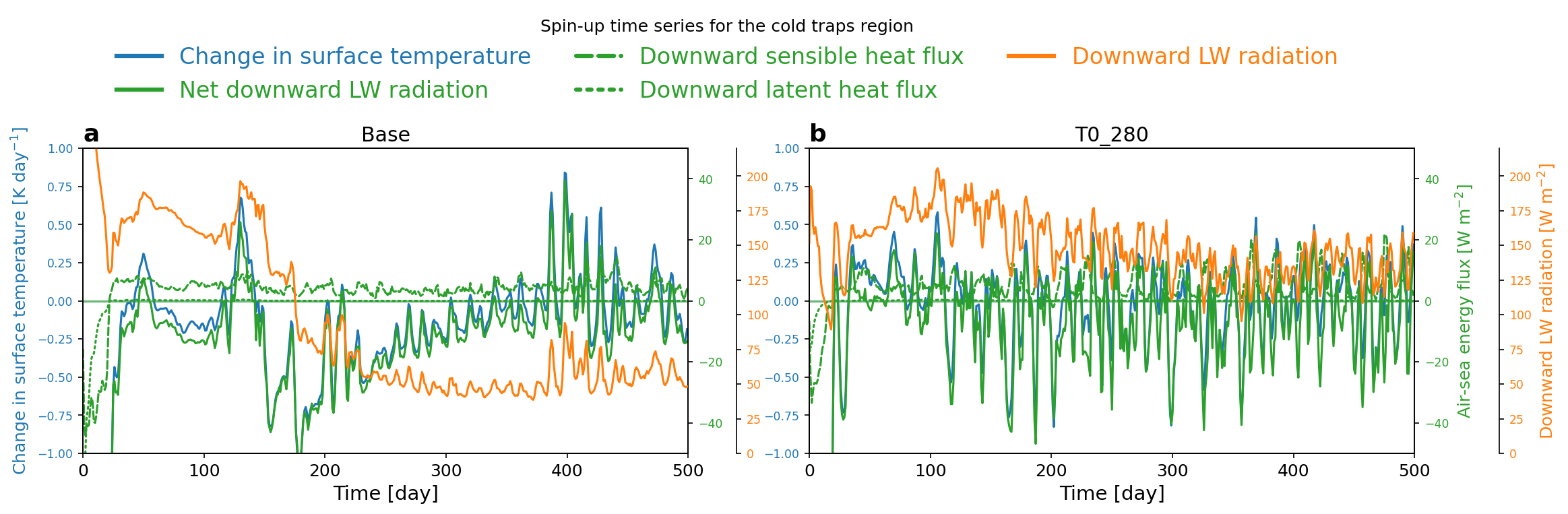}
\caption{Time series of diagnostics for the night-side cold traps, defined as the region bounded by \ang{45} and \ang{55} in the latitude and \ang{160}--\ang{140}W in the longitude. The panels for the (left) SJ and (right) DJ regime show: (orange) downward longwave radiation flux, (green, solid) net longwave radiation flux, (green, dashed) sensible heat flux in \si{\watt\per\m\squared}, (green, dotted) latent heat flux, and (blue) time derivative of the surface temperature in \si{\K\per\day}. \label{fig:cold_traps_fluxes}}
\end{figure*}

The evolution of the atmospheric circulation during the spin-up period causes a substantial decrease of the temperature and humidity on the night side.
Most strikingly, the night-side average surface temperature in the SJ simulation decreases by \SI{40}{\K}, while its minimum temperature drops by almost \SI{60}{\K} (see the blue curve in Fig.~\ref{fig:cold_traps_wvre_cre}).
This change has been noted in our previous work \citep{Sergeev20_atmospheric} and is investigated in more detail in this section, focusing on the initial period of the simulations.
We present the analysis for the night-side cold traps, defined here as the coldest regions of the night side of the planet.
This region is bounded by \ang{45} and \ang{55} in the latitude and \ang{160}--\ang{140}W in the longitude in our simulations.

The time series of the surface temperature in the night-side cold traps aligns well with the time series of global circulation diagnostics such as the wave crest shift and equatorial jet acceleration (cf. Fig.~\ref{fig:crest_lon_eq_jet} and \ref{fig:cold_traps_wvre_cre}).
After the rapid cooling from the initial warm state, the night-side surface reaches \SI{\approx 236}{\K} in both cases.
The temperature in the DJ case further decreases by a few degrees but stays close to this value throughout the simulation (Fig.~\ref{fig:cold_traps_wvre_cre}b).
In the SJ case however, as the circulation regime develops the strong equatorial superrotation and stationary waves (Fig.~\ref{fig:crest_lon_eq_jet}a), the temperature in the cold traps falls by almost \SI{60}{\K} (Fig.~\ref{fig:cold_traps_wvre_cre}a), fluctuating around \SI{\approx 178}{\K} for the remainder of the simulation (see also Fig.~\ref{fig:t_sfc_wvp_cwp}a).

The night-side surface temperature is dictated mostly by the thermal radiation emitted by the atmosphere to the surface, because there is no incident stellar radiation and no dynamic ocean in our setup.
This is confirmed by the time series of energy fluxes shown in Fig.~\ref{fig:cold_traps_fluxes}.
Turbulent heat fluxes are non-zero, but still an order of magnitude smaller than the longwave radiation flux, suppressed by the near-surface temperature inversion \citep{Joshi20_earth}.
Fig.~\ref{fig:cold_traps_fluxes} demonstrates that the downward longwave radiation (orange curve) is a precursor of the surface temperature in the cold traps.
Its substantial decrease (by \SI{> 100}{\watt\per\m\squared}) after about 150 days of the SJ simulation corresponds to the fall in temperature (blue curve in the negative).

The ability of the atmosphere to radiate heat is controlled by its temperature $T_\text{a}$ and emissivity $\epsilon_\text{a}$ \citep[see discussion in e.g.][]{Lewis18_influence}.
The latter is controlled by the amount of water vapor and cloud condensate in the atmosphere, which is shown in Fig.~\ref{fig:cold_traps_wvre_cre} as the water vapor path, i.e. the mass-weighted vertical integral of the water vapor in the atmosphere.
In both SJ and DJ cases, the model simulation starts from a dry state, which is far from the global equilibrium.
This causes an initial spike in the water vapor path (solid purple curves in Fig.~\ref{fig:cold_traps_wvre_cre}).
Subsequently, the water content in the night-side cold traps decreases, dropping in the SJ case to \SI{\approx 0.1}{\kg\per\m\squared}, but remaining an order of magnitude higher in the DJ case, at \SI{\approx 1.1}{\kg\per\m\squared}.
In the SJ case, as the night-side atmosphere becomes drier, it is less able to radiate heat, causing the decrease in the longwave flux received by the surface (orange curve in Fig.~\ref{fig:cold_traps_fluxes}a), which cools as a result.

To show that the effect of water vapor is larger than that of condensed water (clouds), their contributions to the longwave radiative effect near the planet's surface are also plotted in Fig.~\ref{fig:cold_traps_wvre_cre} (red curves).
Following \citet{Eager20_implications}, the radiative effects of water vapor and clouds are isolated using an additional ``diagnostic'' radiative transfer calculation, which does not affect the simulation itself.
On every time step, these additional calculations omit the opacity of water vapor or clouds and are then compared to the ``cloudy'' calculation.
Their difference (``clear-sky'' minus ``cloudy'') is referred to as the cloud radiative effect (CRE).
The more negative the values of the radiative effect in Fig.~\ref{fig:cold_traps_wvre_cre} are, the more important the contribution of the water vapor or clouds is.
Overall, the radiative effect of water vapor is substantially stronger than that of clouds: in the SJ case their time-average values in the second half of the spin-up period are \SI{-28.3}{\watt\per\m\squared} and \SI{-6.7}{\watt\per\m\squared}, respectively; in the DJ case they are \SI{-43.5}{\watt\per\m\squared} and \SI{-30.9}{\watt\per\m\squared}.
The magnitude of the water vapor radiative effect drops substantially in the SJ case, compared to its initial values or those in DJ, which is a direct consequence of the drying of the night-side cold trap regions (and the night side as a whole).
This decrease of the water vapor content on the night side in the SJ simulation is caused by the reduced transport of warm and moist air from the day to the night side.
For the steady state, this was shown previously by \citet{Sergeev20_atmospheric}: the moist static energy flux divergence in the SJ-like regime (their ``\emph{MassFlux}'' case) was smaller than that in the DJ-like regime (their ``\emph{Adjust}'' case).


\subsection{Steady state of the two circulation regimes}
\label{sec:results_mean_state}
\begin{figure*}
    \centering
    \includegraphics[width=\textwidth]{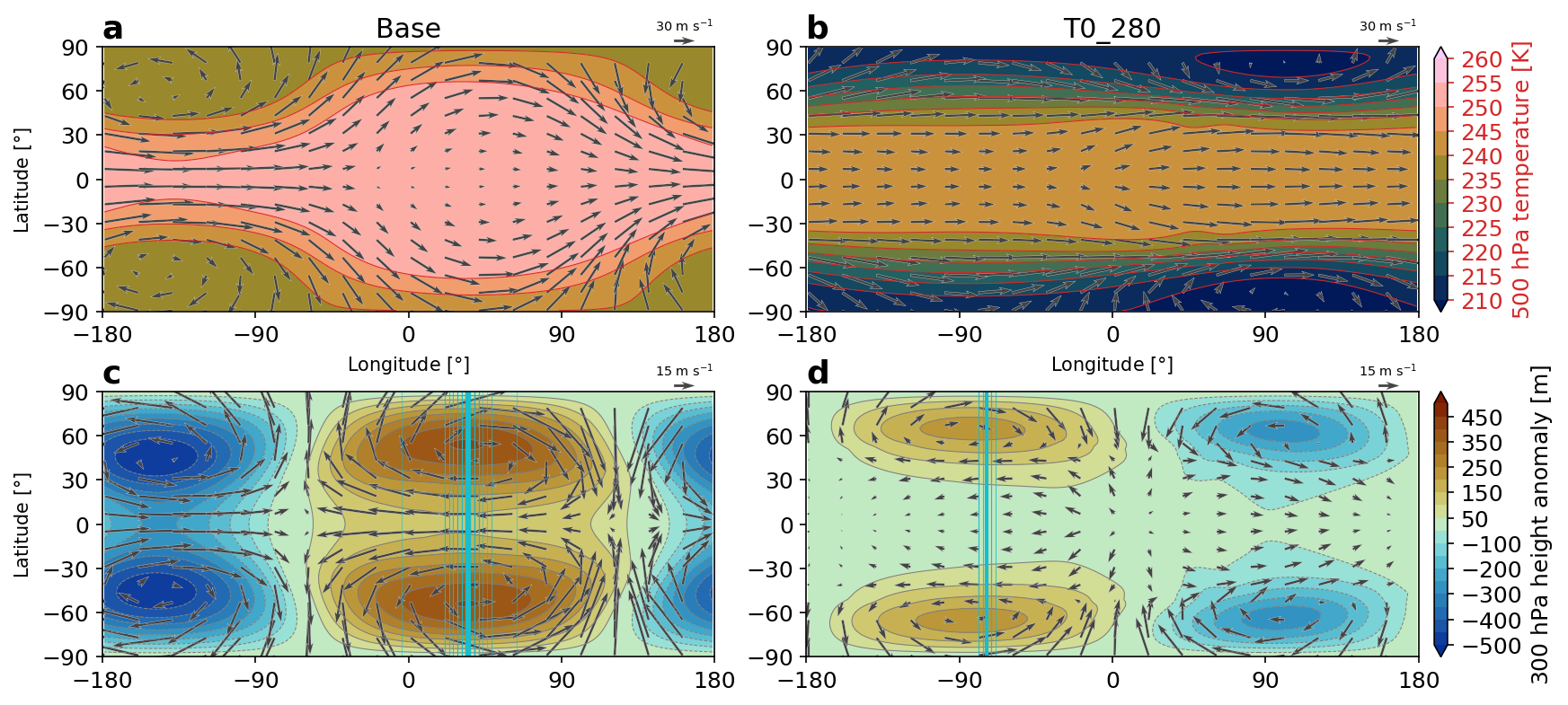}
    \caption{Steady state atmospheric circulation in the (left) \emph{Base} (SJ regime) and (right) \emph{T0\_280} (DJ regime) simulations.
    The panels show (a, b) \SI{500}{\hecto\pascal} temperature (shading, \si{\K}) with \SI{300}{\hecto\pascal} wind vectors, (c, d) \SI{300}{\hecto\pascal} eddy geopotential height (shading, \si{\m}) and eddy wind vectors.
    The cyan lines in the bottom panels show the longitude of the planetary wave crest, defined as the maximum of the geopotential height anomaly.
    Thin cyan lines show the 50-day mean longitude for several time periods of the steady state climate, while the thick cyan line shows the overall time mean longitude.
    The geopotential height anomaly is defined as the deviation from the zonal mean of the height of the \SI{300}{\hecto\pascal} isobaric surface.
    \label{fig:mean_circ}}
\end{figure*}

Focusing on the mature stage of the \emph{Base} and \emph{T0\_280} simulations, we now describe the steady state of the SJ and DJ regimes, respectively.
Model output averaged over the last 1000 days of the simulations is used in this section.
We confirm that the regimes are well-defined and have distinct features in the spatial distribution of the key climate diagnostics: surface and air temperature, total column water vapor and cloud content.
Namely, the SJ regime is characterized by a larger day-night temperature contrast due to extremely cold and dry cloudless regions on the night side, while the DJ regime is characterized by a more zonally-oriented morphology of the wind circulation and temperature, reducing the day-night dichotomy of the planet's climate.
We show that the SJ and DJ regimes are similar to those found by \citet{Edson11_atmospheric} and \citet{Noda17_circulation} with respect to the mean tropospheric conditions.
The upper layers of the atmosphere also have notable differences in variables such as water vapor and cloud content, which has implications for the transmission spectrum of TRAPPIST-1e, as discussed in detail in Sec.~\ref{sec:results_synthobs}.

As summarized for all our simulations in Fig.~\ref{fig:all_sim}c,d, the dominant feature of the global tropospheric circulation in both regimes is prograde (eastward) wind, similar to many previous studies, for both abstract \citep[e.g.][]{Edson11_atmospheric,Carone14_connecting,Haqq-Misra18_demarcating} and specific \citep[e.g.][]{Turbet16_habitability,Boutle17_exploring,Fauchez19_impact} tidally locked planetary configurations.
This is also demonstrated by the maps of wind velocity in Fig.~\ref{fig:mean_circ}a and b.
At the equator, for both regimes the atmosphere is superrotating (local maximum of mean angular momentum, see Eq.~\ref{eq:s_def}), though in the SJ regime it is a dominant feature of the circulation, while in the DJ regime it is weaker than the two eastward jets in the mid-latitudes.

The key difference between the two circulation patterns in their steady state is apparent in the location and amplitude of geopotential height anomalies shown in Fig.~\ref{fig:mean_circ}c,d.
The average longitude of the geopotential maxima represents a wave crest and is marked by the cyan vertical lines (their corresponding time evolution is tracked by the cyan curves in Fig.~\ref{fig:crest_lon_eq_jet}).
In the SJ regime, the geopotential maximum (anticyclone) is to the east of the substellar point, while a pair of minima occupy the night side and correspond to cyclonic gyres (Fig.~\ref{fig:mean_circ}c).
This pattern corresponds to an equatorial Rossby wave, analogous to those generated in Earth's tropics \citep{Vallis20_trouble}, but on a global scale (wavenumber 1) and stationary due to the planet's synchronous rotation \citep{PierrehumbertHammond19_atmospheric}.
This planetary-scale wave pattern is Doppler-shifted eastward by the zonal flow, as discussed in Sec.~\ref{sec:results_spinup_dynamics} \citep[see also][]{ShowmanPolvani11_equatorial}.
The wave is largely geostrophically balanced, as evidenced by Fig.~\ref{fig:mean_circ}c,d because the stationary eddy wind vectors are aligned with geopotential height isolines.
At the equator, the geopotential height in the SJ regime also has a prominent planetary-scale perturbation, which corresponds to an equatorial Kelvin wave \citep[e.g.][]{Debras20_acceleration,WangYang20_phase}.
The superposition of Rossby and Kelvin waves is identical to that obtained in shallow water models of exoplanetary atmospheres without the background flow \citep[e.g.][]{HammondPierrehumbert18_wave-mean,WangYang20_phase}.
The temperature field for the \emph{Base} simulation at high latitudes has a weak gradient from equator to pole, but a strong gradient between the day and night sides (shading in Fig.~\ref{fig:mean_circ}a).
This confirms that the SJ regime is less affected by the extratropical baroclinic instability than the DJ regime (Fig.~\ref{fig:var_tseries}f).

In the DJ regime (the \emph{T0\_280} case), the geopotential height pattern is not shifted by the strong superrotation and so the Rossby wave crest is at the western terminator (\ang{-90} longitude) while its trough straddles the eastern terminator (+\ang{90} longitude, Fig.~\ref{fig:mean_circ}d).
With no wave-jet resonance, the geopotential anomalies are also weaker and located closer to poles, while the height perturbation at the equator is small \citep[see e.g.][]{WangYang20_phase}.
In contrast to the SJ regime, the temperature map is dominated by the meridional gradient instead of the zonal, or day-night, gradient (Fig.~\ref{fig:mean_circ}b, see also the x-axis in Fig.~\ref{fig:all_sim}b).


\begin{figure*}
    \centering
    \includegraphics[width=\textwidth]{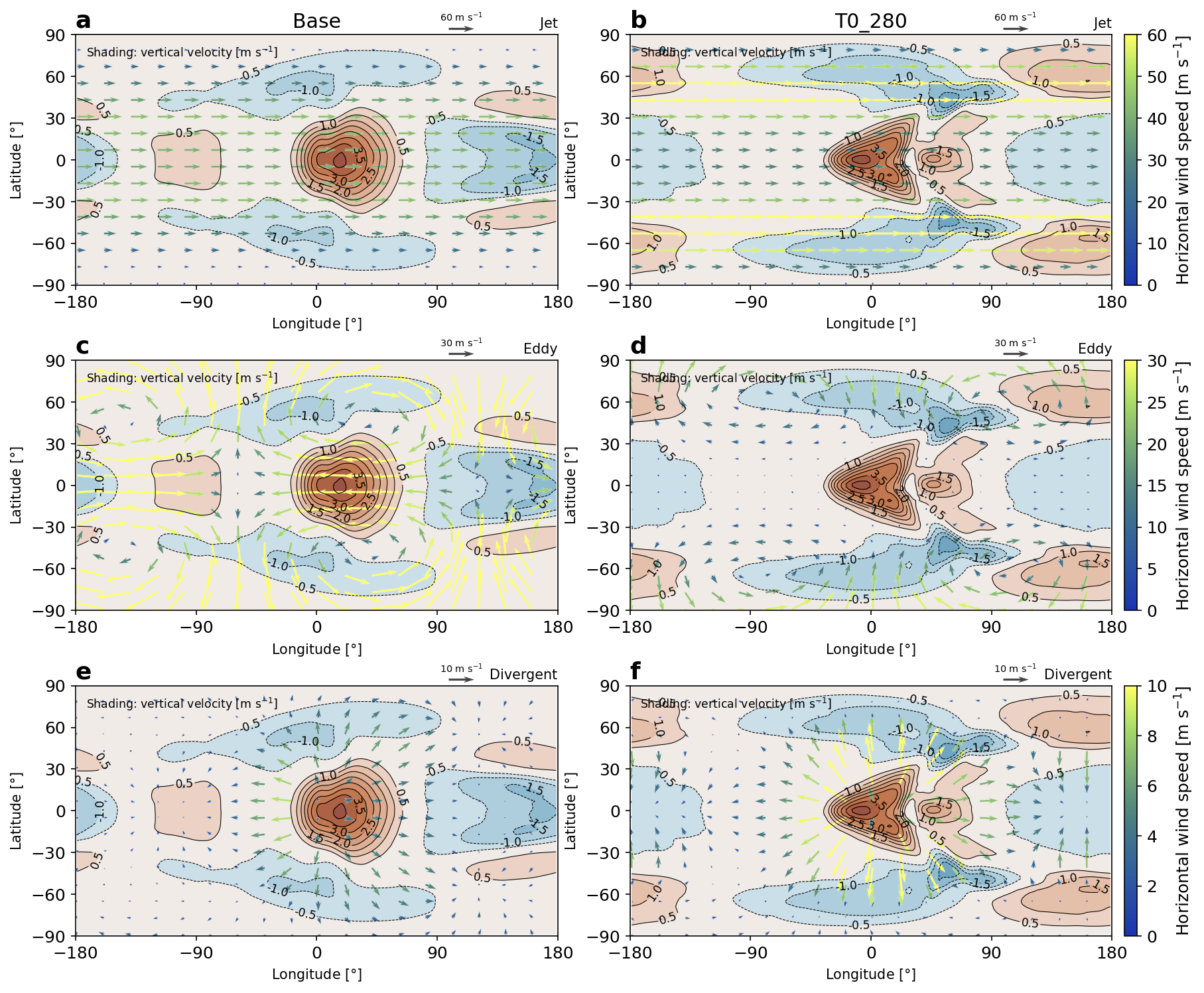}
    \caption{Helmholtz decomposition of the horizontal wind at \SI{300}{\hecto\pascal} in the (left) \emph{Base} (SJ regime) and (right) \emph{T0\_280} (DJ regime) simulations, corresponding to the wind field shown in Fig.~\ref{fig:mean_circ}a,b.
    The panels show (a, b) the zonal mean rotational component, (c, d) the eddy rotational component, (e, f) the divergent component. Note the different scaling of the each of the components.
    Also shown is the upward wind velocity (shading, \si{\m\per\s}).
    \label{fig:rotdiv}}
\end{figure*}

The full 3D structure of the two circulation regimes can be further elucidated by decomposing the wind field into its rotational and divergent components \citep{HammondLewis21_rotational}.
Fig.~\ref{fig:rotdiv} shows this for the \SI{300}{\hecto\pascal} level.
The dominant eastward jets are immediately revealed by taking the zonal average of the rotational flow: a single equatorial jet in the \emph{Base} case and two mid-latitude jets in the \emph{T0\_280} case (Fig.~\ref{fig:rotdiv}a,b).
The eddy component of the rotational flow (Fig.~\ref{fig:rotdiv}c,d) corresponds to the stationary wave pattern (Fig.~\ref{fig:mean_circ}c,d).
The divergent component of the wind flow has a smaller magnitude relative to the rotational wind, but together with the contours of vertical velocity in Fig.~\ref{fig:rotdiv}e,f, it clearly shows the day-night overturning circulation.
The divergent component is notably weaker in the SJ regime (Fig.~\ref{fig:rotdiv}e) and stronger in the DJ regime (Fig.~\ref{fig:rotdiv}f).
The differences in the rotational and divergent components between our simulations are analogous to those found in the THAI results \citep{Turbet22_thai,Sergeev22_thai}, confirming that it is one of the characteristic features of the two regimes.

\begin{figure*}
    \centering
    \includegraphics[width=\textwidth]{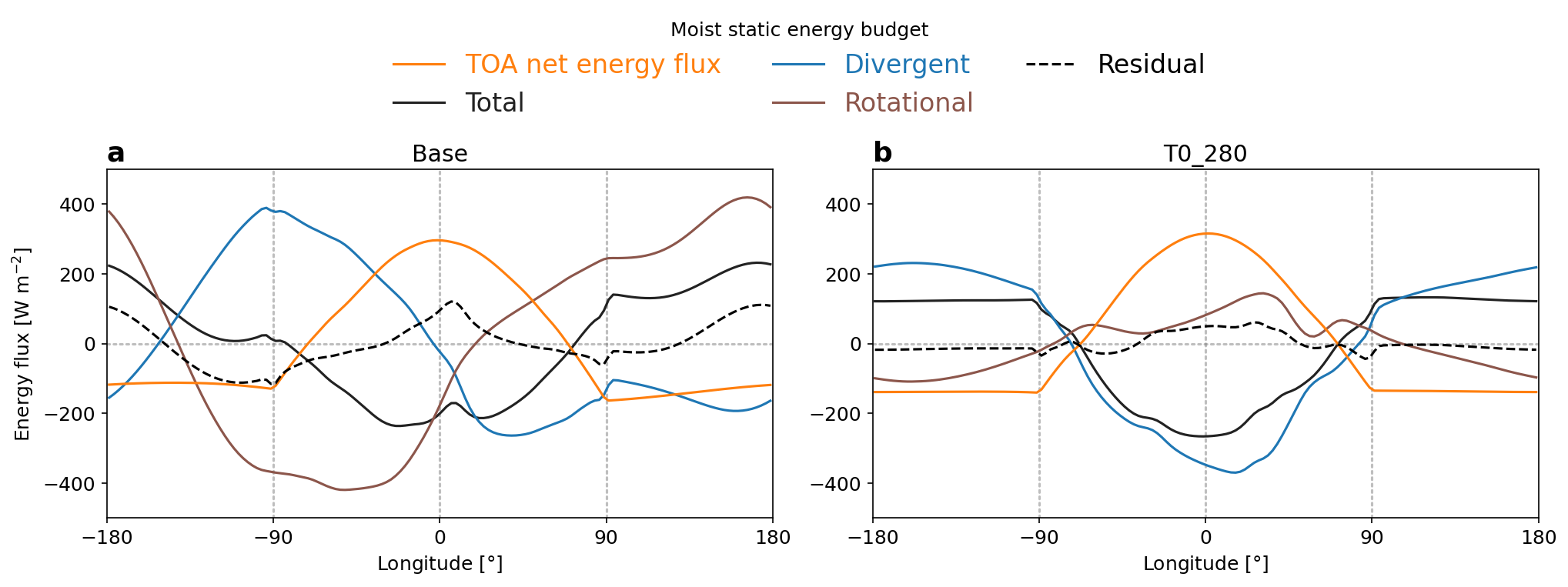}
    \caption{The steady state moist static energy (MSE) budget for the (a) \emph{Base} (SJ regime) and (b) \emph{T0\_280} (DJ regime) simulations.
    See text for details.
    \label{fig:mse_rotdiv}}
\end{figure*}

These differences also result in different relative contributions of the rotational and divergent components of the circulation to the energy transport from the day side to the night side.
We assess this by calculating the moist static energy (MSE) flux divergence for each of the components and show the results in Fig.~\ref{fig:mse_rotdiv}.
MSE is defined as 
\begin{equation}
    h = \underbrace{c_p T + gz}_\text{Dry} + \underbrace{L_v q}_\text{Latent},
\end{equation}
where $c_p$ is the heat capacity at constant pressure, $T$ is temperature, $g$ is the acceleration due to gravity, $z$ is height, $L_v$ is the latent heat of vaporization and $q$ is the water vapor content.
Column-integrated, the divergence of the MSE flux is equal to the total local heating, as expressed by
\begin{equation}
    \langle\nabla\cdot h\vec{u}\rangle + F^{net}_{TOA} = 0,
\end{equation}
where the angle brackets denote a mass-weighted vertical integral, $F^{\mathrm{net}}_{\mathrm{TOA}}$ is the top of the atmosphere net energy flux and $\vec{u}$ is the horizontal wind vector, which can be taken as a rotational or divergent component of the total wind field.

Fig.~\ref{fig:mse_rotdiv}a shows that in the SJ regime the surplus of energy on the day side is redistributed roughly equally by the divergent and rotational components of the flow.
Qualitatively, the divergent component tends to transport MSE from the eastern hemisphere of the planet (to the east of the substellar longitude) to its western hemisphere.
It is balanced by the rotational (jet plus eddy) part, which takes MSE from the western hemisphere and deposits it to the east.
Note that despite the partial cancellation of the jet and eddy components of the rotational flow, its magnitude is still larger than that presented in \citet{HammondLewis21_rotational} for a terrestrial planet case.
This is likely due to the assumption of a weak temperature gradient regime in \citet{HammondLewis21_rotational}, which appears to be less applicable in the \emph{Base} simulation (Fig.~\ref{fig:mse_rotdiv}a).
Another likely reason for the discrepancy is the inter-GCM differences in the boundary layer scheme between our studies and warrants further investigation.
The MSE budget for the DJ regime (Fig.~\ref{fig:mse_rotdiv}b) is similar to that in \citet{HammondLewis21_rotational}, despite the circulation pattern in that study being closer to our SJ regime.
The MSE flux divergence is predominantly due to the divergent component of the flow, while the individual rotational components largely cancel out and make the total rotational MSE flux divergence close to zero.

\begin{figure*}
    \centering
    \includegraphics[width=\textwidth]{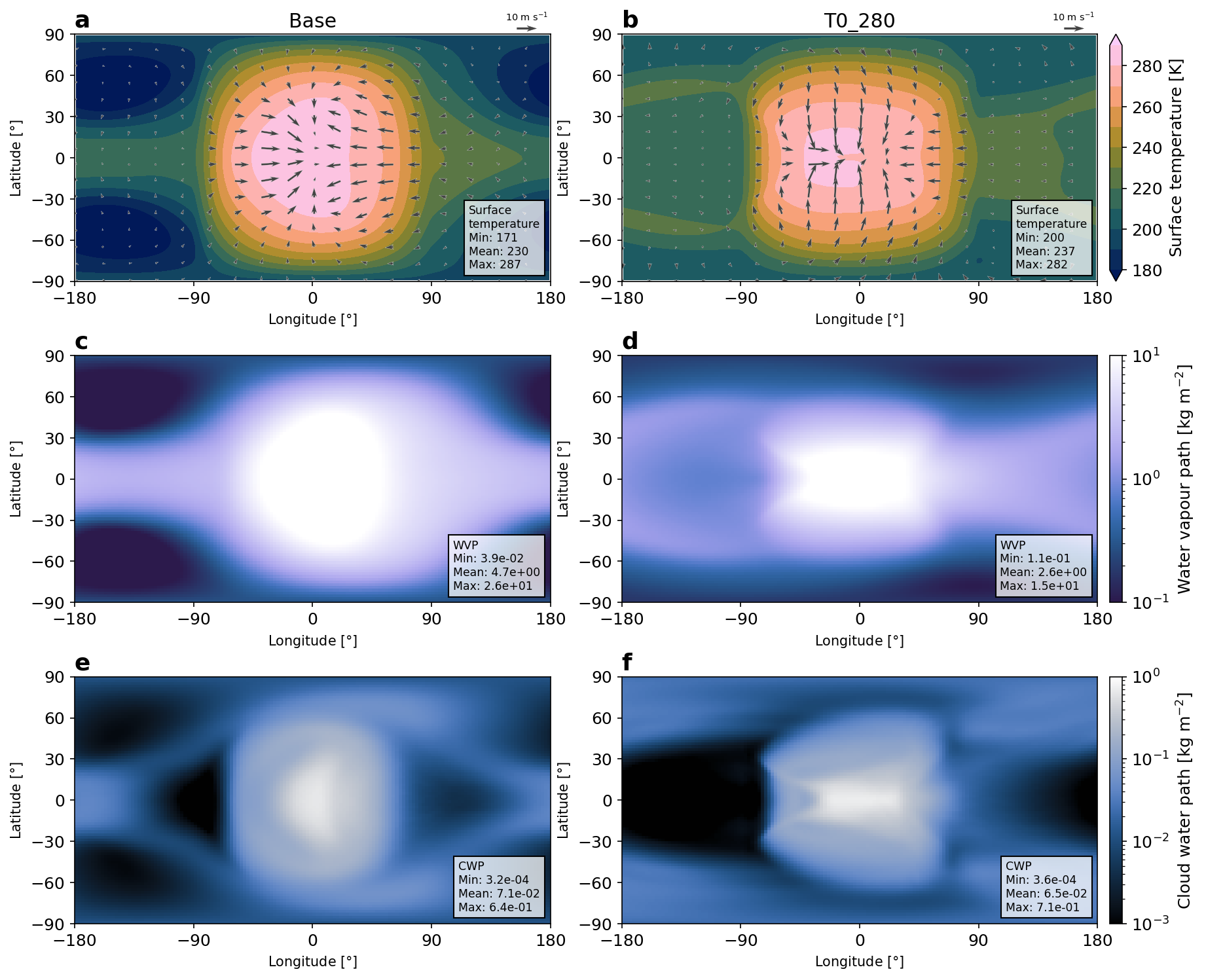}
    \caption{Steady state thermodynamic conditions in the (left) \emph{Base} (SJ regime) and (right) \emph{T0\_280} (DJ regime) simulations.
    The panels show (a, b) surface temperature (shading, \si{\K}) with \SI{10}{\m} wind vectors, (c, d) water vapor path (shading, \si{\kg\per\m\squared}), and (e, f) cloud water path (shading, \si{\kg\per\m\squared}).
    \label{fig:t_sfc_wvp_cwp}}
\end{figure*}

We finish the description of the mean climate by briefly discussing the thermodynamic conditions in the SJ and DJ regimes.
The surface temperature has a spatial distribution similar to that of the mid-tropospheric temperature, with a larger day-night gradient in the SJ regime than in DJ (cf. Fig.~\ref{fig:t_sfc_wvp_cwp}a,b and Fig.~\ref{fig:mean_circ}a,b).
The near surface wind vectors shown in Fig.~\ref{fig:t_sfc_wvp_cwp}a,b demonstrate the region of convergent flow towards the substellar point, which is the lower branch of the overturning circulation shown in Fig.~\ref{fig:rotdiv}e,f.
In the SJ regime, the day side's surface attains a maximum temperature of \SI{287}{\K}, while the minimum temperature is in the night-side cold traps, which are aligned with the cyclonic gyres (Fig.~\ref{fig:mean_circ}c,d) and discussed in more detail in Sec.~\ref{sec:results_spinup_cold_traps}.
In the DJ regime, the surface temperature maximum is about \SI{5}{\K} lower than that in the SJ regime, but the minimum is \SI{29}{\K} higher (Fig.~\ref{fig:t_sfc_wvp_cwp}b).
As a result, the average surface temperatures for the SJ and DJ regime are \SI{230}{\K} and \SI{237}{\K}, respectively.
The surface isotherms and the wind convergence have distinct shapes in each of the regimes: they are broadly oriented zonally in the SJ case and meridionally in the DJ case.
This is also true for the rest of our sensitivity experiments (not shown).

Despite the substantially lower surface temperature minimum in the SJ regime than in the DJ regime, the SJ climate stays temperate and does not reach the condition for a potential atmospheric collapse.
Namely, the lowest temperature remains higher than the \ce{CO2} condensation point for a 1 bar atmosphere, so this species is expected to stay in the gaseous phase throughout the simulation \citep[e.g.][]{Turbet18_modeling}.
The mean surface conditions are such that the substellar region retains the temperature above the freezing point of seawater (Fig.~\ref{fig:t_sfc_wvp_cwp}a,b) and is able to maintain the water cycle on the planet (with the caveat of the globally uniform ocean surface in our setup).
The total area of the ice-free surface (with a temperature above the freezing point of seawater), a crude metric of planetary habitability, is similar for both regimes ($\approx 21$ and \SI{18}{\percent} in the SJ and DJ cases, respectively).
This is close to the estimates by other GCMs in the TRAPPIST-1e simulations with a nitrogen-dominated 1 bar atmosphere \citep[between 20 and \SI{24}{\percent}, see][for more details]{Sergeev22_thai}.

The total column water vapor (water vapor path) broadly mirrors the surface temperature map.
The driest areas clearly match the coldest areas of the surface in the SJ case (Fig.~\ref{fig:t_sfc_wvp_cwp}c), and the day-night asymmetry is overall more pronounced than that in the DJ case (Fig.~\ref{fig:t_sfc_wvp_cwp}d).
The absolute values of water vapor path reach 26 and \SI{15}{\kg\per\m\squared} in the SJ and DJ cases, respectively.
While the SJ case has overall more water vapor in the atmosphere, its driest regions are an order of magnitude drier than those in the DJ case.
Fig.~\ref{fig:t_sfc_wvp_cwp}e,f show the total column cloud condensate (cloud water path, including ice and liquid water).
Its absolute values are rather similar across the two regimes, which is dictated by the same cloud parameterization used in all our simulations --- unlike the inter-regime discrepancy in the THAI Hab~1 simulations \citep[which was due to different parameterizations in different GCMs, see][]{Sergeev22_thai}.
The spatial distribution of the cloud water path is different between the regimes, especially on the night side and at the terminators, which imprints on the transmission spectra (see Sec.~\ref{sec:results_synthobs}).

Our steady state results thus demonstrate that in 3D GCM simulations of TRAPPIST-1e there can exist two well-defined climates with different spatial distribution of winds, temperature, and moisture.
This further confirms one of the major findings of the THAI intercomparison project \citep[e.g.][]{Turbet22_thai}, proving that even with the same planetary setup and even in the same GCM, the circulation can settle in two distinct regimes, SJ and DJ.
Circulation patterns similar to the SJ regime have been reported in previous studies at various degrees of GCM complexity and for various terrestrial atmospheres \citep[e.g.][]{Edson11_atmospheric,Carone15_connecting,Noda17_circulation,Haqq-Misra18_demarcating}.
For example, it resembles the circulation obtained at an intermediate range of planetary rotation rate in \citet{Edson11_atmospheric}.
Later it was also found in the idealized experiments of \citet{Noda17_circulation}, who labeled this regime as ``Type II''.
In their setup, this regime developed at the rotation period roughly between 5 and 20 Earth days --- comparable to the rotation rate of TRAPPIST-1e (Table~\ref{tab:planet}).
\citet{Noda17_circulation} likewise attribute the emergence of this regime to the resonant excitation of the planetary-scale stationary waves seen in our \emph{Base} simulation.
Our SJ regime also corresponds to the ``Rhines rotator'' circulation regime in \citet{Haqq-Misra18_demarcating}.

The DJ regime was also identified by \citet{Edson11_atmospheric}, \citet{Noda17_circulation}, and \citet{Haqq-Misra18_demarcating}, in the experiments with the planetary rotation period smaller than 1--4 Earth days.
\citet{Noda17_circulation} labeled this circulation pattern as ``Type IV'' and identified the flow features highly similar to those in the \emph{T0\_280} case here.
The similarity extends even to the precipitation field (as a proxy for convective activity), oriented more zonally than that in the SJ regime (Fig.~\ref{fig:t_sfc_wvp_cwp}e,f).

The key difference between the studies mentioned above and our study is that they typically define different circulation regimes by varying planetary or stellar parameters, such as the planet's rotation rate, over a large range of values; while our study focuses on one specific exoplanet.
The regime bistability in our simulations could be further investigated using 3D GCMs for example by running a model ensemble with slightly different initial conditions or with perturbed parameters in sub-grid parameterizations.
It is also pertinent to extend our study to other exoplanetary atmospheres that may be susceptible to regime bistability explored here for TRAPPIST-1e.
This should help to narrow observational constraints of atmospheric dynamics on rocky exoplanets in general.

\subsection{Implications for observations}
\label{sec:results_synthobs}
\subsubsection{Terminator-mean transmission spectra}
\begin{figure*}
\includegraphics[width=\textwidth]{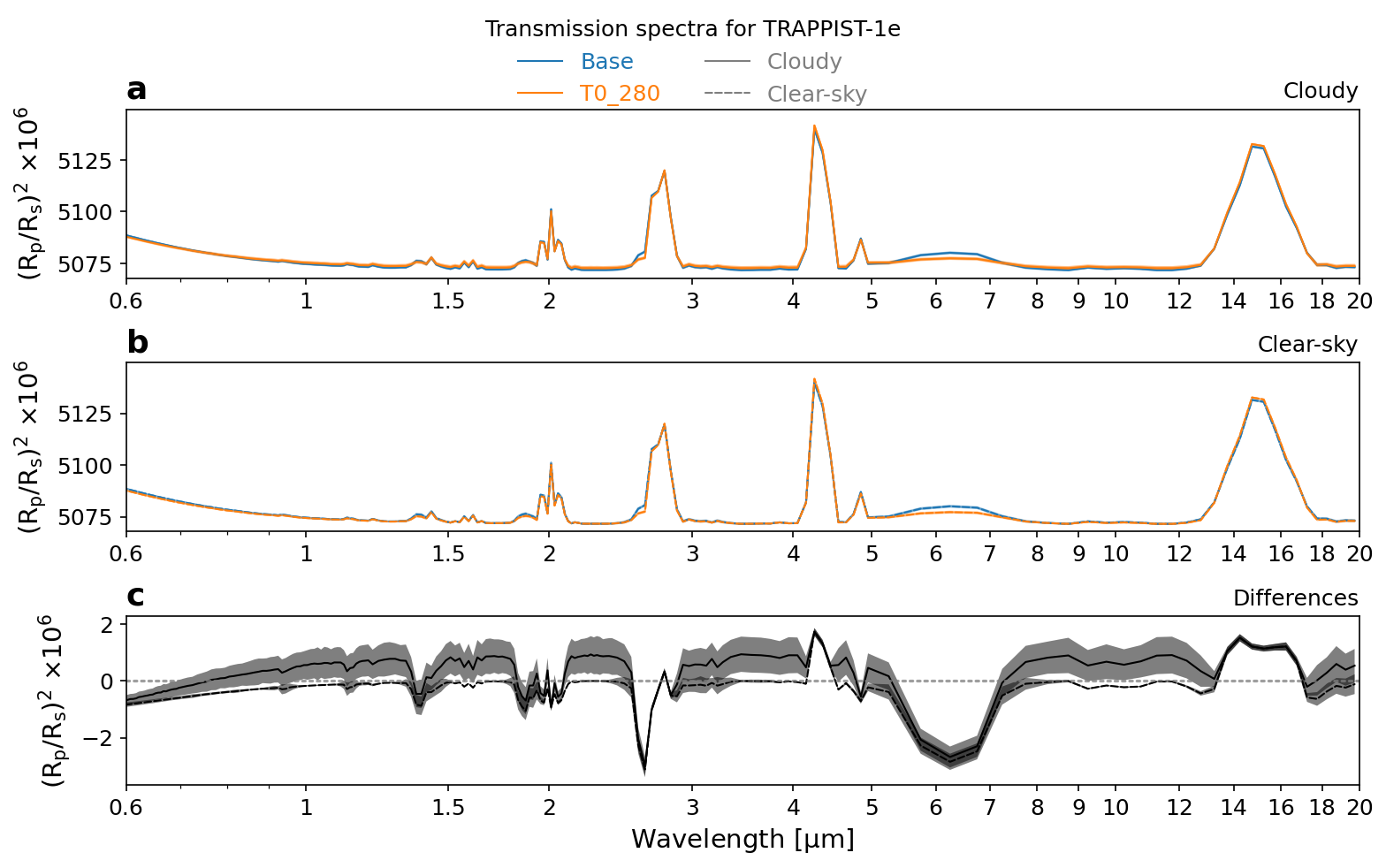}
\caption{Simulated transmission spectra for the (blue) SJ and (orange) DJ cases. The transmission spectra are calculated using fluxes (a) with and (b) without the effect of clouds included. Panel c shows the difference, SJ minus DJ, for (solid lines) cloudy and (dashed lines) clear-sky calculations. \label{fig:synthobs}}
\end{figure*}
Atmospheric characterization of transiting terrestrial exoplanets is becoming feasible with powerful new observational facilities such as the James Webb Space Telescope (JWST), successfully launched in December 2021.
TRAPPIST-1e is the most promising target known so far for such studies \citep{Fauchez19_impact,Suissa20_dim}, mostly thanks to the short orbital period of the planet and the small size of its host star, an ultra-cool M8V dwarf.
It has been shown that inter-model differences, namely in the amount of clouds at the terminator, affect the number of transits required for a confident detection of atmospheric features \citep{Fauchez22_thai}.
Here, we test if the distinct circulation regimes with their water vapor and cloud differences have a detectable imprint in a synthetic transmission spectrum.
We use the same two simulations as before, \emph{Base} and \emph{T0\_280}, corresponding to the SJ and DJ regimes, respectively.

Synthetic transmission spectra are computed natively within the radiation scheme of the UM (SOCRATES), once a day over 61 Earth days (10 orbits) during the steady state phase of the simulation (Sec.~\ref{sec:method_synthobs}).
The time mean and typical time variability ($\pm$ standard deviation) for both SJ and DJ regimes are shown in Fig.~\ref{fig:synthobs}.
The most prominent peaks correspond to the \ce{CO2} absorption bands at 2.7, 4.3 and \SI{15}{\micro\m}.
These peaks are a robust feature of our simulations and are largely unaffected by the presence of clouds, as the difference curve shows in Fig.~\ref{fig:synthobs}c.
The same result has been obtained in other 3D GCMs \citep{Fauchez22_thai} and is explained by the fact that even above the cloud deck there is enough \ce{CO2} to saturate the absorption lines.

While small relative to absorption peaks, differences between the SJ and DJ regimes are consistent in the continuum level, which is higher in DJ case across most of the spectrum between 0.6 and \SI{20}{\micro\m} (Fig.~\ref{fig:synthobs}c).
To explain this, we plot the time-mean vertical profiles of cloud content at the terminator in Fig.~\ref{fig:term_diff_climate}.
The profiles reveal that for the DJ case clouds tend to occur at lower altitudes than in the SJ case.
However, the mixing ratio of cloud ice (whose content dominates over cloud water) is noticeably larger in DJ case compared to the SJ case (Fig.~\ref{fig:term_diff_climate}d), and this leads to a slightly higher continuum level for the former, by \SI{\approx 1}{ppm}.

The water vapor band at \SI{6.3}{\micro\m}, on the other hand, is stronger by up to \SI{2.5}{ppm} in SJ than in the DJ case.
This can be attributed to a much lower water vapor content in the upper layers in the DJ regime compared to the SJ regime, as demonstrated by the vertical profiles in Fig.~\ref{fig:term_diff_climate}b.
This difference is similar to that between the LMD-G model and three other GCMs in the THAI project \citep{Fauchez22_thai}.
Note however, that the LMD-G model exhibits an SJ-like circulation regime in the THAI Hab~1 simulation --- so its low humidity in the upper atmosphere is likely a consequence of using a convection adjustment scheme \citep{Sergeev22_thai}.

\subsubsection{East-west terminator differences}
\label{sec:term_asymmetry}
\begin{figure*}
\includegraphics[width=\textwidth]{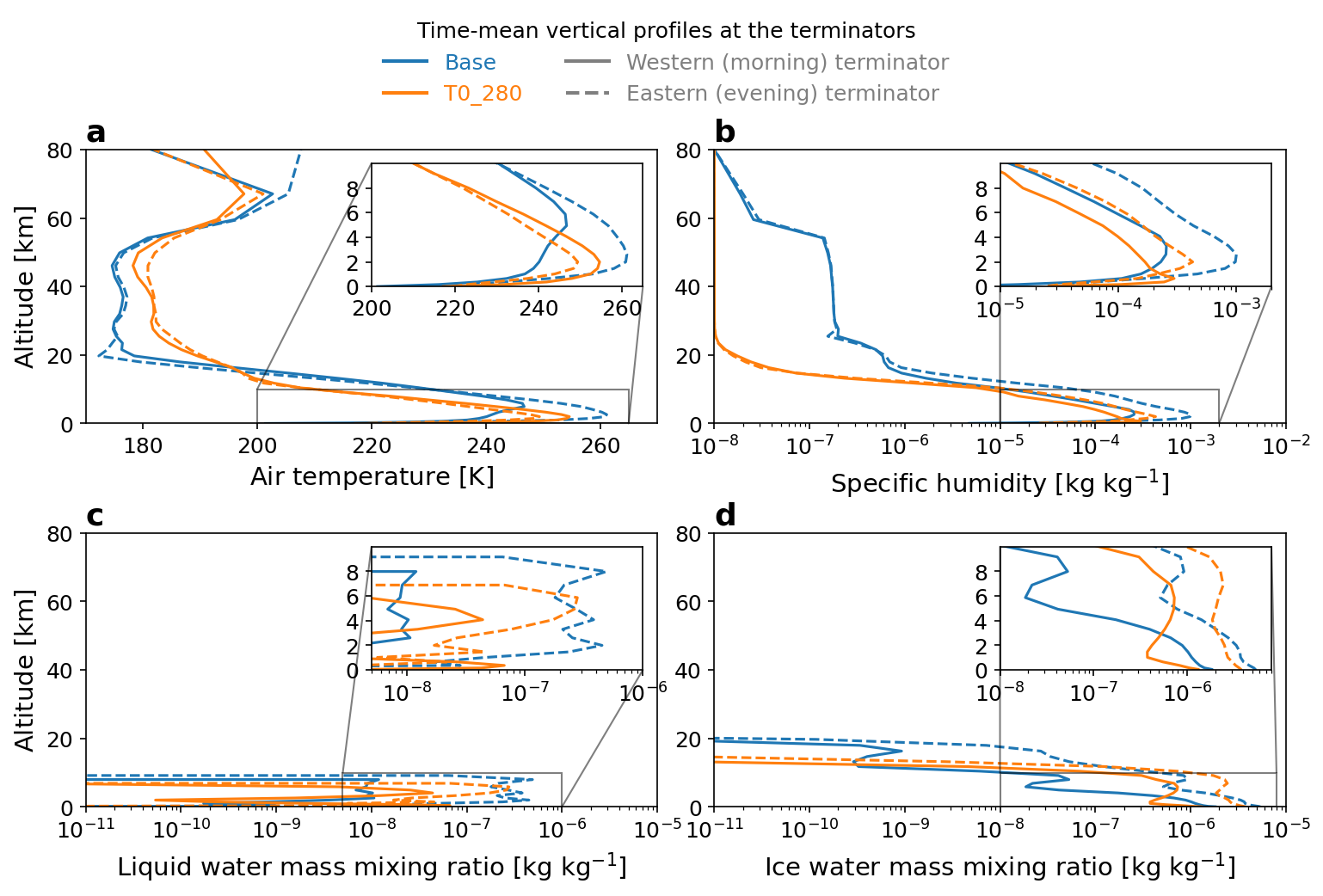}
\caption{Time mean vertical profiles at the (solid lines) western and (dashed lines) eastern terminators in the (blue) SJ and (orange) DJ. The variables shown are (a) air temperature, (b) specific humidity, (c) cloud liquid water mixing ratio, (d) cloud ice mixing ratio. \label{fig:term_diff_climate}}
\end{figure*}
Even though our simulations assume a uniform ocean surface covering the whole planet, conditions at the western and eastern terminators are not the same.
This is due to the zonal asymmetry in the global circulation introduced by a superposition of the stationary eddies and mean overturning circulation (see Sec.~\ref{sec:results_spinup_dynamics}).
It is important to take the asymmetry into account, because averaging the transmission spectrum over the full terminator may cancel out absorption features.
The terminator asymmetry in the transmission spectra was found to be non-negligible in previous studies, both for hotter gas giants \citep[e.g.][]{Line16_influence,Powell19_transit} and colder rocky planets \citep{SongYang21_asymmetry}.
Here we confirm that the circulation regime differences result in slightly different transmission spectra at the eastern and western terminators.
We find that the terminator asymmetry is roughly twice as large for the SJ regime compared to that in the DJ regime.

\begin{figure*}
\includegraphics[width=\textwidth]{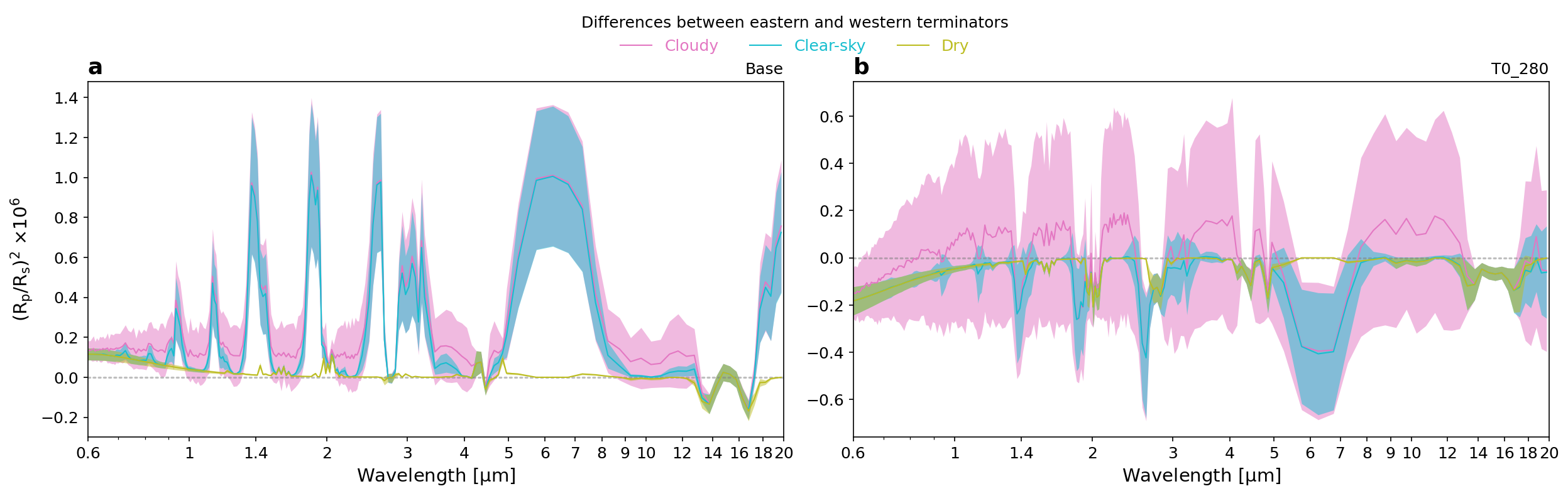}
\caption{Differences in the transmission depth (ppm) between terminators (eastern minus western) in the (a) SJ and (b) DJ cases. Different lines show spectral differences assuming (pink) cloudy, (cyan) clear-sky, and (olive) dry atmosphere.\label{fig:term_diff_spectra}}
\end{figure*}

To determine what contributes to the asymmetry the most, we obtain differences in the spectra using all-sky (i.e. cloudy), clear-sky (i.e. cloudless), and dry (i.e. excluding the opacity of water vapor) radiation calculations, which are shown separately for the two regimes in Fig.~\ref{fig:term_diff_spectra}a and b, respectively.
The bulk of terminator asymmetry in both cases is due to water vapor at its absorption bands (e.g. at $\approx$1.4, 1.9, 2.7 and \SI{6.3}{\micro\m}).
This is confirmed by the curves corresponding to cloudy and clear-sky calculations being close together in these regions (compare the pink and cyan curves in Fig.~\ref{fig:term_diff_spectra}).
Between the \ce{H2O} absorption bands, terminator asymmetry is of the order of \SI{\approx 0.25}{ppm}, showing that the eastern terminator is cloudier and thus slightly elevates the continuum level.
The eastern terminator has more clouds, mostly in the form of ice crystals, than its western counterpart (compare solid and dashed lines in Fig.~\ref{fig:term_diff_climate}c,d).
Temperature asymmetry between terminators is different in the SJ compared to DJ case (Fig.~\ref{fig:term_diff_climate}a), but it contributes very little to the asymmetry in transmission depth, being only somewhat visible in the \ce{CO2} absorption bands, as well as in the Rayleigh scattering slope at the shorter wavelengths.

In the SJ regime, the terminator differences (eastern minus western) are overall mostly positive, which is similar to the fast-rotating simulation in \citet{SongYang21_asymmetry}, but have a much lower magnitude of about \SIrange{1}{2}{ppm} --- closer to the slow-rotating simulation in the same study.
The east-west terminator difference has the same sign as that found for cloudy simulations of hot Jupiters, but the magnitude is several orders smaller \citep[e.g.][]{Powell19_transit}, because the atmospheres of rocky planets (as assumed in our study) are much thinner than that of gas giants.
In the DJ regime, the asymmetry is less pronounced, because the overall circulation is more zonally symmetric (Fig.~\ref{fig:mean_circ}) and the differences between terminators are muted (Fig.~\ref{fig:term_diff_climate}b).
The most notable inter-regime difference in the transmission spectra is in the water absorption regions, the largest of which is centered at \SI{6.3}{\micro\m}.
This terminator asymmetry at \SI{6.3}{\micro\m} is positive in the SJ regime, meaning there is more water vapor at the eastern than at the western terminator (Fig.~\ref{fig:term_diff_spectra}a), but in DJ regime the opposite is true.
Note this is difficult to see in the mean vertical profiles in Fig.~\ref{fig:term_diff_climate}b, because the absolute and relative values of water vapor are very small in the upper atmosphere.

The differences in transmission spectra between our simulations, as well as the terminator asymmetry, are too small to be observable with instruments aboard the JWST \citep{May21_water}.
For example, a recent study by \citet{Rustamkulov22_analysis} reports 14 and \SI{10}{ppm} noise floors with 3-$\sigma$ and 1.7-$\sigma$ confidence levels, respectively, for JWST's near infrared spectrograph (NIRSpec) instrument.
The transmission depth differences in our study are even smaller than those simulated for a planet like TRAPPIST-1e by \citet{SongYang21_asymmetry}.
This is because ExoCAM used in that study tends to have higher cloud decks at the terminators compared to other commonly used GCMs, while the UM tends to have lower clouds \citep{Fauchez22_thai}.
Furthermore, ExoCAM was shown to produce consistently moister atmospheres than the UM for simulations of TRAPPIST-1e with an Earth-like atmospheric composition \citep{Sergeev22_thai,Wolf22_exocam}, thus amplifying the potential zonal asymmetry in transmission spectra.

Future telescopes may yet be precise enough to reveal the differences between circulation regimes, which could be elucidated by looking at transmission spectra for both individual and averaged terminator data.
More studies are needed for better understanding of the cloud microphysics affecting the atmospheric opacity on exoplanets like TRAPPIST-1e.

\section{Conclusions}
We investigated the bistability of the atmospheric circulation in the climate simulations of TRAPPIST-1e assuming aquaplanet surface conditions and an \ce{N2}-dominated \SI{1}{bar} moist atmosphere.
The key findings of this study are as follows.
\begin{enumerate}
    \item The emerging atmospheric circulation can have two distinct regimes, either dominated by a strong equatorial eastward jet (the SJ regime) or by a pair of mid-latitude eastward jets (the DJ regime).
    The SJ and DJ regimes correspond to the ``Type II'' and ``Type IV'' regimes in \citet{Noda17_circulation}, or to ``Rhines rotator'' and ``fast rotator'' in the terminology of \citet{Haqq-Misra18_demarcating}.
    The states are well defined and there are practically no intermediate regimes, with respect to key climate diagnostics (e.g. the day-night temperature gradient).
    \item In our simulations which of the regimes the climate enters is sensitive to several factors: a change in physical parameterization, a different surface boundary condition for the temperature, or different initial conditions.
    However, the regime bistability is not merely an artifact of our GCM: similar circulation regimes were recently identified using other GCMs in both dry \citep{Turbet22_thai} and moist \citep{Sergeev22_thai} simulations of TRAPPIST-1e climate; as well as in earlier idealized studies \citep[e.g.][]{Edson11_atmospheric}.
    An interesting outcome of our sensitivity simulations is the bistability due the initial conditions and specifically the initial temperature.
    Namely, at certain moderate initial temperatures the DJ regime develops instead of the SJ regime as in the control simulation (started from \SI{300}{\K}).
    This finding complements the studies by \citet{Thrastarson10_effects} and \citet{Cho15_sensitivity} who used a 3D GCM of a hot Jupiter with a simplified representation of boundary-layer friction and thermal forcing.
    They found that the steady-state atmospheric circulation is sensitive to the initial conditions, unless a much stronger momentum and thermal drag is applied at the bottom of the atmosphere, thus damping any small-scale variability as was done in \citet{Liu13_atmospheric}.
    We leave the sensitivity of the atmospheric circulation on TRAPPIST-1e to the initial wind profile for future studies. 
    \item Using two indicative simulations, one started from \SI{300}{\K} and another from \SI{280}{\K}, we analyze the evolution of the SJ and DJ regimes, respectively.
    We show that the initial stage of the regime evolution depends on a fine balance between the zonally asymmetric heating due to the planet's synchronous rotation on the one hand and mean overturning circulation on the other.
    The SJ regime appears to be weakly radiatively forced at the beginning due to a higher concentration of water vapor and thus stronger longwave cooling of the atmosphere compared to that in the early stages of the DJ regime.
    Consequently, the nascent equatorial jet accelerates more gradually in the SJ case and is able to achieve resonance with the stationary Rossby wave, which in turn markedly amplifies, reinforcing the jet.
    The DJ regime appears to be relatively strongly forced because of the lower initial temperature and thus lower water vapor concentration in the first tens of days; thus the nascent equatorial jet accelerates at a higher rate.
    However, after about 80 days the circulation transitions to the two mid-latitude jets (i.e. the DJ pattern), which are associated with colder polar regions and thus a higher degree of baroclinicity.
    As a result, the wave-jet resonance is not achieved.
    \item The zonal angular momentum budget further supports these arguments.
    In the initial stage of the regime evolution, the surplus of the angular momentum at the equator (i.e. superrotation) is provided by the stationary eddies, which are initially stronger in the DJ case.
    At about day 80, the mean advection terms in the DJ case grow, deplete the zonal momentum at the equator and move it poleward, starting to actively accelerate the mid-latitude jets.
    Meanwhile, the SJ regime matures via the stationary eddy contribution to the angular momentum budget at the equator, which intensifies as the stationary Rossby wave pattern resonates with the jet.
    \item Having fully developed, the two circulation regimes each have a slightly different imprint on the transmission spectrum, though the differences are too small to be observable with the current technology.
    The DJ regime has more clouds at the terminators, so its continuum level is higher than that for SJ over the most part of the analyzed wavelength range (\SIrange{0.6}{20}{\micro\m}).
    The upper atmosphere water vapor content, on the other hand, is higher in the SJ regime, so the \ce{H2O} absorption, especially near \SI{6.3}{\micro\m} is higher.
    Comparable to the inter-regime differences, there is also an asymmetry between eastern and western terminators, which is more pronounced in the SJ regime.
\end{enumerate}

It is clear from both this study that TRAPPIST-1e resides in a particularly sensitive position with respect to the circulation regime, and even small changes in the model setup can tip the circulation into one regime or another.
This exoplanet is one of the key targets for the upcoming JWST observational programs \citep{Gillon20_trappist1}, so understanding its atmospheric structure is imperative for the best use of observations.
We expect that our results could be applicable to other rocky exoplanets residing in a similar ``sweet spot'' of the planetary size and rotation period.
As indicated by earlier modeling studies of hypothetical planets, there are transition regions between well-defined circulation regimes for which a similar regime bistability and sensitivity to GCM setup can exist (\citealp{Edson11_atmospheric}; \citealp{Noda17_circulation}; see also Fig.~1 in \citealp{Carone18_stratosphere}).
Atmospheric circulation on TRAPPIST-1e appears to be particularly sensitive --- not only to the model choice \citep{Turbet22_thai,Sergeev22_thai}, but also to a small change in the initial conditions (this study).
Furthermore, different regimes can emerge not only for a nitrogen-dominated atmosphere, but for a \ce{CO2} atmosphere too, as noted in the THAI project.
With regards to atmospheric pressure, our preliminary experiments with the total pressure below 1~bar favor the SJ regime, while those with the pressure above 1~bar tend to favor the DJ regime, though a separate study is needed for a confident conclusion.
We have also conducted a series of dry simulations starting the UM from different initial temperatures, i.e. repeating the \emph{T0\_250}, \emph{T0\_260}, \emph{T0\_270}, \emph{T0\_280}, \emph{T0\_290}, and \emph{Base} simulations with a dry atmosphere.
The atmospheric circulation in all of these runs evolves into only one regime, namely SJ.
This result indicates that the bistability is driven primarily by moisture effects, at least in our model.
However, we would like to stress that different states can be obtained for the same planet (and the same initial setup) using different GCMs even in dry conditions \citep{Turbet22_thai}, which is important, even if it is not strictly a bifurcation.
Further work on the underlying dynamical mechanisms of the emergence of the two regimes is required, both for dry and moist setups.

While our study explores the emergence of the circulation regimes in depth and across a few sensitivity experiments, a wider modeling study is needed to outline what GCM configurations favor SJ or DJ regimes.
This has been initiated by the THAI project \citep{Fauchez20_thai_protocol}, but would benefit from expanding the parameter sweep wider, e.g. to other configurations with a non-Earth atmospheric composition.
Additionally, our recommendation for modeling atmospheres prone to regime bistability is to use initial condition and/or physical parameterization ensembles.

The surface boundary conditions are among the factors the regime is sensitive to in our simulations.
As shown by \citet{Lewis18_influence} and \citet{Salazar20_effect}, the presence of a continent on the day side of the planet affects the global circulation.
Such a perturbation to the model may favor one of the circulation regimes, depending on the size and thermodynamic properties of the continent.
Including a dynamic ocean will likely influence the regime bistability too by contributing to the heat transport between the day and night sides of the planet \citep[see e.g.][]{Hu14_role,DelGenio19_habitable}.
These avenues of research are left for the future.

Finally, our simulations are performed with time invariant stellar forcing.
The high sensitivity of the circulation regimes even to initial conditions indicates that the circulation may be prone to an abrupt transition if a temporary forcing is provided.
Such a forcing can be an influx of water vapor into the stratosphere in the aftermath of a large volcanic eruption \citep[e.g.][]{Loffler16_impact,Guzewich22_volcanic} or a series of eruptions \citep[e.g.][]{Joshi03_gcm}.
Another example is periodic change in stellar forcing due to flaring of the host star, which as an M-dwarf.
Performing experiments with periodic or transient stellar forcing \citep[as in e.g.][]{Chen21_persistence} may further elucidate the question of bistability of atmospheric circulation on TRAPPIST-1e.

\acknowledgments
The authors would like to thank two anonymous reviewers for their constructive comments, which helped to improve the manuscript.
Material produced using Met Office Software.
We acknowledge use of the Monsoon2 system, a collaborative facility supplied under the Joint Weather and Climate Research Programme, a strategic partnership between the Met Office and the Natural Environment Research Council.
This work was supported by a Science and Technology Facilities Council Consolidated Grant (\texttt{ST/R000395/1}), UKRI Future Leaders Fellowship (\texttt{MR/T040866/1}), and the Leverhulme Trust (\texttt{RPG-2020-82}).
NTL was supported by Science and Technology Facilities Council Grant \texttt{ST/S505638/1}.
IAB and JM acknowledge support of the Met Office Academic Partnership secondment.

\software{
The Met Office Unified Model is available for use under license; see \url{http://www.metoffice.gov.uk/research/modelling-systems/unified-model}.
Scripts to post-process and visualize the model data are available on GitHub: \url{https://github.com/dennissergeev/t1e_bistability_code} and are dependent on the following open-source Python libraries: \texttt{aeolus} \citep{aeolus}, \texttt{cmcrameri} \citep{Crameri20_misuse}, \texttt{iris} \citep{iris}, \texttt{jupyter} \citep{jupyter}, \texttt{matplotlib} \citep{matplotlib}, \texttt{numpy} \citep{numpy}, \texttt{windspharm} \citep{windspharm}.
          }

\appendix
\section{Stationary wave pattern and zonal mean atmospheric structure during the model spin-up}
\label{app:spinup}
\begin{figure*}
    \centering
    \includegraphics[width=\textwidth]{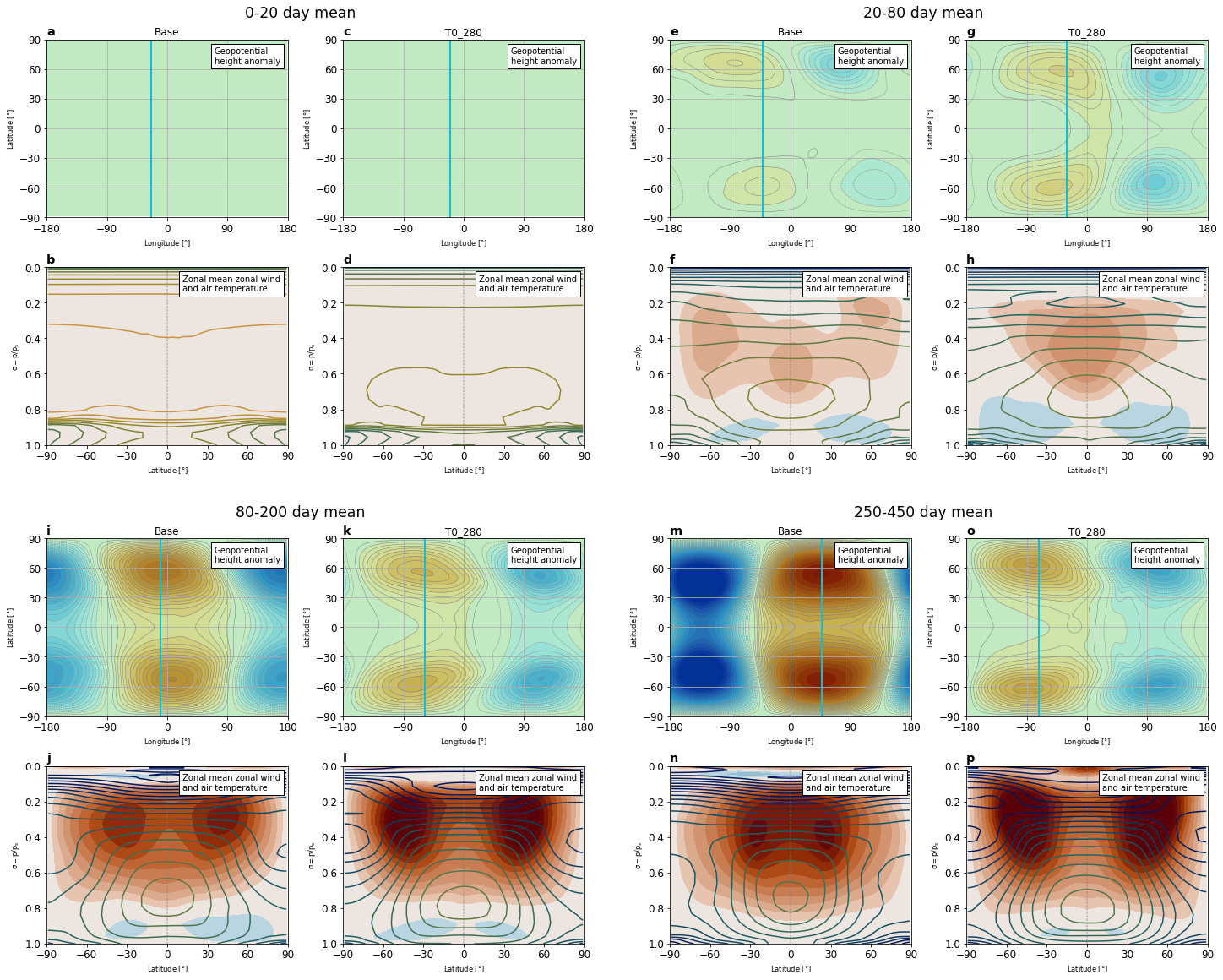}
    \caption{This figure corresponds to the animation provided as Supplemental Video 1 and shows the stationary wave pattern and zonal mean atmospheric structure during the four time periods discussed in Sec.~\ref{sec:results_spinup_dynamics}.
    Fig.~\ref{fig:all_sim}c,d and \ref{fig:mean_circ}c,d show the corresponding variables in steady state.
    Panels (a, c, e, g, i, k, m, o) show the eddy geopotential height defined as the deviation from the zonal mean of the height of the \SI{300}{\hecto\pascal} isobaric surface (shading, \si{\m}).
    The cyan lines show the mean longitude of the planetary wave crest, defined as the maximum of the geopotential height anomaly.
    Panels (b, d, f, h, j, l, n, p) show the vertical cross-section of the zonal mean eastward wind (shading, \si{\m\per\s}) and air temperature (contours, \si{\K}).
    All variables are averaged over the periods of (a, b, c, d) 0--20 days, (e, f, g, h) 20--80 days, (i, j, k, l) 80--200 days, and (m, n, o, p) 250--450 days.}
    \label{fig:spinup_ghgt_anom_wind_temp}
\end{figure*}
\clearpage

\section{Derivation of the angular momentum budget}
\label{app:aam}
We present here the derivation of the angular momentum budget equation given by Eq.~\eqref{eq:aam_main}, starting from \citet[][Sec.~2.2.7]{Vallis17_aofd}.
The zonal component of the axial angular momentum per unit mass is 
\begin{equation}
    m = (u + \Omega r\cos\phi)r\cos\phi,
\end{equation}
where $u$ is the zonal component of the wind velocity vector, $\Omega$ is the rotation rate, $r$ is the planetary radius, and $\phi$ is latitude.
The conservation equation for this quantity is analogous to the zonal momentum equation and in spherical coordinates may be written as
\begin{equation}
    \frac{Dm}{Dt} = -\frac{1}{\rho}\frac{\partial p}{\partial\lambda} - G_\lambda,
    \label{eq:aam_start}
\end{equation}
where $D/Dt$ is the material derivative, $\rho$ is density, $\lambda$ is longitude, and $G_\lambda$ represents friction and dissipation.
The material derivative in spherical coordinates is
\begin{equation}
    \frac{D}{Dt} = \frac{\partial}{\partial t} + \frac{u}{r\cos\phi}\frac{\partial}{\partial\lambda} + \frac{v}{r}\frac{\partial}{\partial\phi} + w\frac{\partial}{\partial r},
\end{equation}
where $t$ is time, $v$ and $w$ are the meridional and vertical components of the wind velocity vector, respectively.
Using the mass continuity equation
\begin{equation}
    \frac{\partial\rho}{\partial t} = \frac{1}{r\cos\phi}\frac{\partial\rho u}{\partial\lambda} + \frac{1}{r\cos\phi}\frac{\partial}{\partial\phi}(\rho v\cos\phi) + \frac{1}{r^{2}}\frac{\partial}{\partial r}(\rho wr^{2}) = 0,
    \label{eq:cont_full}
\end{equation}
and expanding the material derivative, \eqref{eq:aam_start} can be written as
\begin{equation}
    \frac{\partial\rho m}{\partial t}+\frac{1}{r\cos\phi}\frac{\partial(\rho um)}{\partial\lambda} + \frac{1}{r\cos\phi}\frac{\partial}{\partial\phi}(\rho vm\cos\phi)+\frac{1}{r^{2}}\frac{\partial}{\partial r}(\rho wmr^{2})=-\frac{\partial p}{\partial\lambda}-r\cos\phi\rho G_{\lambda}.
    \label{eq:aam_cons}
\end{equation}

We then average \eqref{eq:aam_cons} over time and longitude, to provide a budget for the zonal mean angular momentum with separate terms associated with contributions from the mean flow, stationary eddies and transient eddies.
First, we define the corresponding averaging operations for a quantity $X$ as
\begin{align}
    \overline{X}&=\frac{1}{\Delta T}\int^{\Delta T}_0X\,\mathrm{d}t, \label{eq:t_avg}\\ 
    [X] &= \frac{1}{2\pi}\int^{2\pi}_0X\,\mathrm{d}\lambda. \label{eq:z_avg} 
\end{align}
\eqref{eq:t_avg} and \eqref{eq:z_avg} represent the time and zonal average operations, respectively.
Subtracting the averages, we get the corresponding ``eddy'' quantities
\begin{align}
    X^{\prime} &= X - \overline{X}, \\ 
    X^{\ast} &= X - [X],
\end{align}
which represent transient and stationary eddies, respectively.
We can then re-write the instantaneous quantity as 
\begin{equation}
    X = [\overline{X}] + \overline{X}^{\ast} +X^{\prime}.
\end{equation}
Following \citet[][Sec.~4.1]{PeixotoOort92_physics}, we also note that the time and zonal product of two quantities $X$ and $Y$ is given by
\begin{equation}
    [\overline{XY}] = [\overline{X}][\overline{Y}] + [\overline{X}^{\ast}\overline{Y}^{\ast}] + [\overline{X^{\prime}Y^{\prime}}]. \label{eq:full_split}
\end{equation}

We now apply the time and zonal averaging operations, \eqref{eq:t_avg} and \eqref{eq:z_avg}, respectively, to \eqref{eq:aam_cons}.
The integration over longitude in the zonal average means that terms involving $\partial/\partial\lambda$ become zero.
Time and zonal averaging gives us
\begin{equation}
    \frac{\Delta[\rho m]}{\Delta T} + \frac{1}{r\cos\phi}\frac{\partial}{\partial\phi}([\overline{\rho vm}]\cos\phi) + \frac{1}{r^{2}}\frac{\partial}{\partial r}([\overline{\rho wm}]r^{2})= - r\cos\phi[\overline{\rho G_\lambda}],
    \label{eq:tz_avg}
\end{equation}
where $\Delta[\rho m] = ([\rho m]_{t=T} - [\rho m]_{t=0})$.
When the atmospheric circulation is in a steady state, this term approaches zero. 

Using \eqref{eq:full_split}, the second and third terms in \eqref{eq:tz_avg} can be decomposed into transport terms associated with the mean meridional flow, the stationary eddies, and the transient eddies.
Substituting $V=\rho v$ and $W=\rho w$ for brevity, we obtain
\begin{align}
    \frac{\Delta[\rho m]}{\Delta T} +\frac{1}{r\cos\phi}\frac{\partial}{\partial\phi}([\overline{V}][\overline{m}]\cos\phi) + \frac{1}{r^{2}}\frac{\partial}{\partial r}([\overline{W}][\overline{m}]r^{2}) &= -\frac{1}{r\cos\phi}\frac{\partial}{\partial\phi}([\overline{V}^{\ast}\overline{m}^{\ast}]\cos\phi) - \frac{1}{r^{2}}\frac{\partial}{\partial r}([\overline{W}^{\ast}\overline{m}^{\ast}]r^{2}) \nonumber \\ 
    &\phantom{=\ \,} -\frac{1}{r\cos\phi}\frac{\partial}{\partial\phi}([\overline{V^{\prime}m^{\prime}}]\cos\phi) - \frac{1}{r^{2}}\frac{\partial}{\partial r}([\overline{W^{\prime}m^{\prime}}]r^{2}) \nonumber \\
    &\phantom{=\ \,} - r\cos\phi[\overline{\rho G_\lambda}].
    \label{eq:aam_tz_avg}
\end{align}
Note that for simplicity the time derivative term and dissipation term are not expanded. 

The right-hand-side of \eqref{eq:aam_tz_avg} can be re-written as an advection of mean angular momentum by making use of the continuity equation \eqref{eq:cont_full}.
After time and zonal averaging, the continuity equation becomes
\begin{equation}
    \frac{\Delta[\rho]}{\Delta T} +\frac{1}{r\cos\phi}\frac{\partial}{\partial\phi}([\overline{V}]\cos\phi) + \frac{1}{r^{2}}\frac{\partial}{\partial r}([\overline{W}]r^{2}) = 0 \label{eq:cont}
\end{equation}
where again we have re-written $V=\rho v$ and $W = \rho w$, and the $\Delta$ symbol has the same meaning as before. 

Upon multiplication by $[\overline{m}]$, \ref{eq:cont} can be re-arranged to yield
\begin{equation}
    [\overline{m}]\frac{\Delta[\rho]}{\Delta T}+\frac{1}{r\cos\phi}\frac{\partial}{\partial\phi}([\overline{V}][\overline{m}]\cos\phi)+\frac{1}{r^{2}}\frac{\partial}{\partial r}([\overline{W}][\overline{m}]r^{2}) - \frac{[\overline{V}]\cos\phi}{r\cos\phi}\frac{\partial[\overline{m}]}{\partial\phi} - \frac{[\overline{W}]r^{2}}{r^{2}}\frac{\partial[\overline{m}]}{\partial r} = 0,
\end{equation}
or, canceling the $\cos\phi$ and $r$ terms,
\begin{equation}
    \frac{1}{r\cos\phi}\frac{\partial}{\partial\phi}([\overline{V}][\overline{m}]\cos\phi)+\frac{1}{r^{2}}\frac{\partial}{\partial r}([\overline{W}][\overline{m}]r^{2}) = \frac{[\overline{V}]}{r}\frac{\partial[\overline{m}]}{\partial\phi} + [\overline{W}]\frac{\partial[\overline{m}]}{\partial r} - [\overline{m}]\frac{\Delta[\rho]}{\Delta T}
    \label{eq:funky_cont}
\end{equation}
Substitution of \eqref{eq:funky_cont} into \eqref{eq:aam_tz_avg} and ignoring the change of zonal mean density with time yields
\begin{align}
    \frac{\Delta[\rho m]}{\Delta T} &=
    - \frac{[\overline{V}]}{r}\frac{\partial[\overline{m}]}{\partial\phi} - [\overline{W}]\frac{\partial[\overline{m}]}{\partial r} \nonumber\\
    &\phantom{=\ \,} -\frac{1}{r\cos\phi}\frac{\partial}{\partial\phi}([\overline{V}^{\ast}\overline{m}^{\ast}]\cos\phi) - \frac{1}{r^{2}}\frac{\partial}{\partial r}([\overline{W}^{\ast}\overline{m}^{\ast}]r^{2}) \nonumber\\ 
    &\phantom{=\ \,} -\frac{1}{r\cos\phi}\frac{\partial}{\partial\phi}([\overline{V^{\prime}m^{\prime}}]\cos\phi) - \frac{1}{r^{2}}\frac{\partial}{\partial r}([\overline{W^{\prime}m^{\prime}}]r^{2}) \nonumber\\
    &\phantom{=\ \,} - r\cos\phi[\overline{\rho G_\lambda}].
\end{align}
This equation is equivalent to \eqref{eq:aam_main} used in Sec.~\ref{sec:results_spinup_dynamics}.

Additionally, it can be shown that in a steady state in the absence of friction and dissipation, zonal angular momentum is materially conserved by the mean flow.
We do this by dropping the time derivative term (first term in the previous equation) and substituting $[\overline{\mathrm{D}}]/\mathrm{D}[\overline{t}]=([\overline{V}]/r)\partial/\partial\phi + [\overline{W}]\partial/\partial r$, which yields
\begin{align}
    \frac{[\overline{\mathrm{D}}][\overline{m}]}{\mathrm{D}[\overline{t}]} &= -\frac{1}{r\cos\phi}\frac{\partial}{\partial\phi}([\overline{V}^{\ast}\overline{m}^{\ast}]\cos\phi) - \frac{1}{r^{2}}\frac{\partial}{\partial r}([\overline{W}^{\ast}\overline{m}^{\ast}]r^{2}) \nonumber\\
    &\phantom{=\ \,} -\frac{1}{r\cos\phi}\frac{\partial}{\partial\phi}([\overline{V^{\prime}m^{\prime}}]\cos\phi) - \frac{1}{r^{2}}\frac{\partial}{\partial r}([\overline{W^{\prime}m^{\prime}}]r^{2}).
\end{align}
We can also see that if dissipation transports $[\overline{m}]$ down gradient, and friction acts to restore $[\overline{m}]$ toward a state of solid body co-rotation with the underlying planet, then eddies that can transport $[\overline{m}]$ up gradient are required to maintain a local maximum of $[\overline{m}]$ such that
\begin{equation}
    s=\frac{[\overline{m}]}{\Omega r^{2}}-1>0,
    \label{eq:s_def}
\end{equation}
where $s>0$ indicates superrotation \citep[see e.g.][]{Read18_superrotation,Lewis21_characterizing}.

\bibliography{references}{}
\bibliographystyle{aasjournal}
\end{document}